\pdfoutput=1

\documentclass[11pt,twoside,a4paper,cmspaper,final,collab]{cms-tdr}
\def\svnVersion{256823}\def\cmsCernNo{2014-051}\def\cmsCernDate{\today}\def\cmsMessage{Published in the European Physical Journal C as \href{http://dx.doi.org/10.1140/epjc/s10052-014-2980-6}{\doi{10.1140/epjc/s10052-014-2980-6.}}}

\begin{document}\cmsNoteHeader{HIG-13-030}

\hyphenation{had-ron-i-za-tion}
\hyphenation{cal-or-i-me-ter}
\hyphenation{de-vices}

\RCS$Revision: 254481 $
\RCS$HeadURL: svn+ssh://svn.cern.ch/reps/tdr2/papers/HIG-13-030/trunk/HIG-13-030.tex $
\RCS$Id: HIG-13-030.tex 254481 2014-08-05 10:54:09Z jbrooke $
\newcommand{\AddJetMaxCSV}{\ensuremath{\mathrm{CSV}_{\mathrm{aj}}}}
\newcommand{\AddJetMindR}{\ensuremath{\Delta R(\PH,\mathrm{aj})}}
\newcommand{\BRinv}{\ensuremath{\mathcal{B}(\PH\to\text{inv})}}
\newcommand{\CLs}{\ensuremath{\mathrm{CL}_\mathrm{s}}\xspace}
\newcommand{\delphill}{\ensuremath{\Delta\phi_{\ell\ell}}}
\newcommand{\dphiMJ}{\ensuremath{\Delta\phi(\MET,\mathrm{j})}}
\newcommand{\dphiMtkM}{\ensuremath{\Delta\phi(\MET,\MET{}_\text{trk})}}
\newcommand{\dphiZH}{\ensuremath{\Delta\phi(\cPZ,\PH)}}
\newcommand{\dRJJ}{\ensuremath{\Delta R_\mathrm{jj}}}
\newcommand{\dThPull}{\ensuremath{\Delta\theta_{\text{pull}}}}
\newcommand{\dyll}{\ensuremath{\cPZ/\gamma^*\to\ell^+\ell^-}}
\newcommand{\etajj}{\ensuremath{\Delta \eta_\mathrm{jj}}}
\newcommand{\GamInv}{\ensuremath{\Gamma_{\text{inv}}}}
\newcommand{\Hinv}{\ensuremath{\PH(\text{inv})}}
\newcommand{\mH}{\ensuremath{m_{\PH}}}
\newcommand{\mjj}{\ensuremath{M_\mathrm{jj}}}
\newcommand{\mll}{\ensuremath{m_{\ell\ell}}}
\newcommand{\mt}{\ensuremath{m_\mathrm{T}}}
\newcommand{\mW}{\ensuremath{m_{\PW}}}
\newcommand{\mZ}{\ensuremath{m_{\cPZ}}}
\newcommand{\NA}{\text{---}}
\newcommand{\Naj}{\ensuremath{N_{\mathrm{aj}}}}
\newcommand{\phijj}{\ensuremath{\Delta \phi_\mathrm{jj}}}
\newcommand{\pta}{\ensuremath{\pt^\mathrm{j1}}}
\newcommand{\ptb}{\ensuremath{\pt^\mathrm{j2}}}
\newcommand{\ptjj}{\ensuremath{\pt^\mathrm{jj}}}
\newcommand{\ptV}{\ensuremath{\pt(\mathrm{V})}}
\newcommand{\VHbb}{\ensuremath{\mathrm{V}\PH(\bbbar)}}
\newcommand{\VJ}{\ensuremath{\mathrm{V}\text{+jets}}}
\newcommand{\Wb}{\ensuremath{\PW\mathrm{+}\cPqb}}
\newcommand{\Wbb}{\ensuremath{\PW\mathrm{+}\bbbar}}
\newcommand{\Wjets}{\ensuremath{\PW\text{+jets}}\xspace}
\newcommand{\WlnH}{\ensuremath{\PW(\ell\cPgn)\PH(\bbbar)}}
\newcommand{\Wt}{\ensuremath{\PW\cPqt}}
\newcommand{\Wudscg}{\ensuremath{\PW\mathrm{+udscg}}}
\newcommand{\WZ}{\ensuremath{\PW\cPZ}}
\newcommand{\Zb}{\ensuremath{\cPZ\mathrm{+}\cPqb}}
\newcommand{\Zbb}{\ensuremath{\cPZ\mathrm{+}\bbbar}}
\newcommand{\ZbbH}{\ensuremath{\cPZ(\bbbar)\PH}}
\newcommand{\ZbbHinv}{\ensuremath{\cPZ(\bbbar)\PH(\text{inv})}}
\newcommand{\Zjets}{\ensuremath{\cPZ\text{+jets}}\xspace}
\newcommand{\ZllH}{\ensuremath{\cPZ(\ell\ell)\PH}}
\newcommand{\ZllHinv}{\ensuremath{\cPZ(\ell\ell)\PH(\text{inv})}}
\newcommand{\ZnnH}{\ensuremath{\cPZ(\cPgn\cPagn)\PH(\bbbar)}}
\newcommand{\Ztobb}{\ensuremath{\cPZ\to\bbbar}}
\newcommand{\Zudscg}{\ensuremath{\cPZ\mathrm{+udscg}}}
\newcommand{\ZZ}{\ensuremath{\cPZ\cPZ}}
\newlength\cmsFigWidth
\ifthenelse{\boolean{cms@external}}{\setlength\cmsFigWidth{0.95\columnwidth}}{\setlength\cmsFigWidth{0.8\textwidth}}
\ifthenelse{\boolean{cms@external}}{\providecommand{\cmsLeft}{top}}{\providecommand{\cmsLeft}{left}}
\ifthenelse{\boolean{cms@external}}{\providecommand{\cmsRight}{bottom}}{\providecommand{\cmsRight}{right}}
\newcommand{\tauh}{\ensuremath{\Pgt_\mathrm{h}}\xspace}

\cmsNoteHeader{HIG-13-030} 
\title{Search for invisible decays of Higgs bosons in the vector boson fusion and associated ZH production modes}

\date{\today}

\abstract{A search for invisible decays of Higgs bosons is performed using the vector boson fusion and associated ZH production modes.  In the ZH mode, the Z boson is required to decay to a pair of charged leptons or a $\bbbar$ quark pair.  The searches use the 8\TeV pp collision dataset collected by the CMS detector at the LHC, corresponding to an integrated luminosity of up to 19.7\fbinv. Certain channels include data from 7\TeV collisions corresponding to an integrated luminosity of 4.9\fbinv. The searches are sensitive to non-standard-model invisible decays of the recently observed Higgs boson, as well as additional Higgs bosons with similar production modes and large invisible branching fractions.  In all channels, the observed data are consistent with the expected standard model backgrounds.  Limits are set on the production cross section times invisible branching fraction, as a function of the Higgs boson mass, for the vector boson fusion and ZH production modes.  By combining all channels, and assuming standard model Higgs boson cross sections and acceptances, the observed (expected) upper limit on the invisible branching fraction at $m_\PH=125$\GeV is found to be 0.58\,(0.44) at 95\% confidence level.  We interpret this limit in terms of a Higgs-portal model of dark matter interactions.}

\hypersetup{%
pdfauthor={CMS Collaboration},%
pdftitle={Search for invisible decays of Higgs bosons in the vector boson fusion and associated ZH production modes},%
pdfsubject={CMS},%
pdfkeywords={CMS, physics, Higgs}}

\maketitle 

\section{Introduction}

The discovery of a Higgs boson~\cite{Aad:2012tfa,Chatrchyan:2012ufa,Chatrchyan:2013lba}, together with the absence of any experimental hint of physics beyond the standard model (SM) at the Large Hadron Collider (LHC), have had a major impact on proposed theoretical models for new physics.  All measurements of the recently observed 125\GeV boson to date indicate compatibility with the SM Higgs boson, but the associated uncertainties are large, and the possibility for non-SM properties remains.  Moreover, although additional SM-like Higgs bosons have been excluded over a wide range of masses, additional Higgs bosons with exotic decay modes remain a possibility.

Invisible Higgs boson decays are possible in a wide range of models, for example through decays to neutralinos in supersymmetric models~\cite{Belanger:2001am}, or graviscalars in models with extra dimensions~\cite{Giudice:2000av,Battaglia:2004js}.  In general, interactions of the Higgs boson with the unknown dark matter (DM) sector may introduce invisible decay modes, and bounds on these decays can constrain DM models.  In so-called ``Higgs-portal'' models of DM interactions~\cite{Patt:2006fw,Djouadi:2011aa,Djouadi:2012zc}, the Higgs boson takes the role of mediator between the SM particles and the DM particle.  Recent theories proposing that the Higgs boson played a central role in the evolution of the early universe~\cite{Servant:2013uwa} provide further motivation to understand the relationship between the Higgs boson and DM.

Indirect constraints on non-SM decay modes of the recently observed Higgs boson have been inferred from the visible SM decay modes by including an additional non-SM partial width term in the combined fit to the data~\cite{Chatrchyan:2013lba}.  The resulting upper limit on the non-SM branching fraction is 0.89, at 95\% confidence level (CL).  Direct searches for invisible Higgs boson decays, \Hinv, are possible by requiring that the Higgs boson recoils against a visible system.  Such searches were performed at LEP~\cite{Searches:2001ab,Abdallah:2003ry,Abbiendi:2006gd}, using the ZH associated production mode.  They excluded at 95\% CL an invisible Higgs boson of mass smaller than 105\GeV and produced with a cross section higher than 0.2 times the standard model ZH cross section. Phenomenological studies of hadron collider searches for \Hinv\ have considered all production mechanisms~\cite{Choudhury:1993hv,Martin:1999qf,Eboli:2000ze,Godbole:2003it,Davoudiasl:2004aj,Bai:2011wz,Ghosh:2012ep}. Recently, the ATLAS Collaboration reported a search for invisible decays of a Higgs boson produced in association with a Z boson that decays to leptons~\cite{Aad:2014iia}, placing an upper limit on the invisible Higgs boson branching fraction of 0.75 at 95\% CL for $\mH=125.5$\GeV. The ATLAS Collaboration also searched for an invisibly decaying Higgs boson in association with either a W or Z boson decaying hadronically~\cite{Aad:2013oja}.

\begin{figure*}[ht]
	\begin{center}
		 \includegraphics[width=0.3\textwidth]{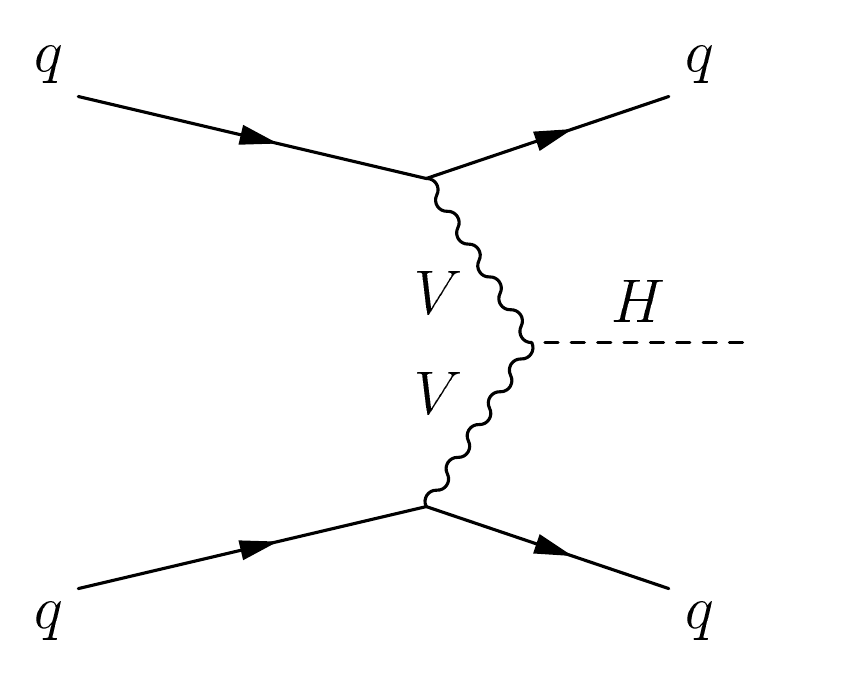}
		 \includegraphics[width=0.3\textwidth]{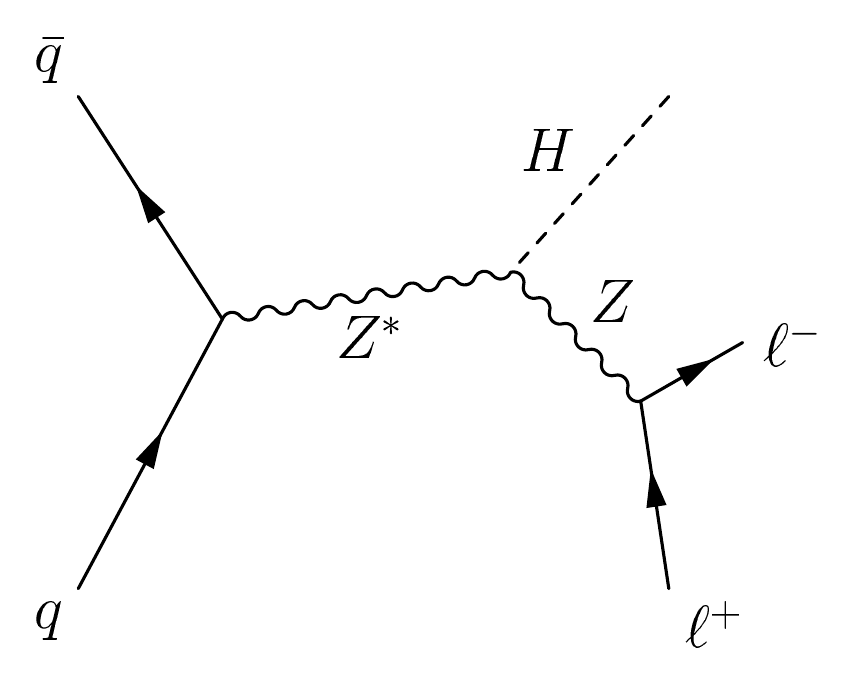}
		 \includegraphics[width=0.3\textwidth]{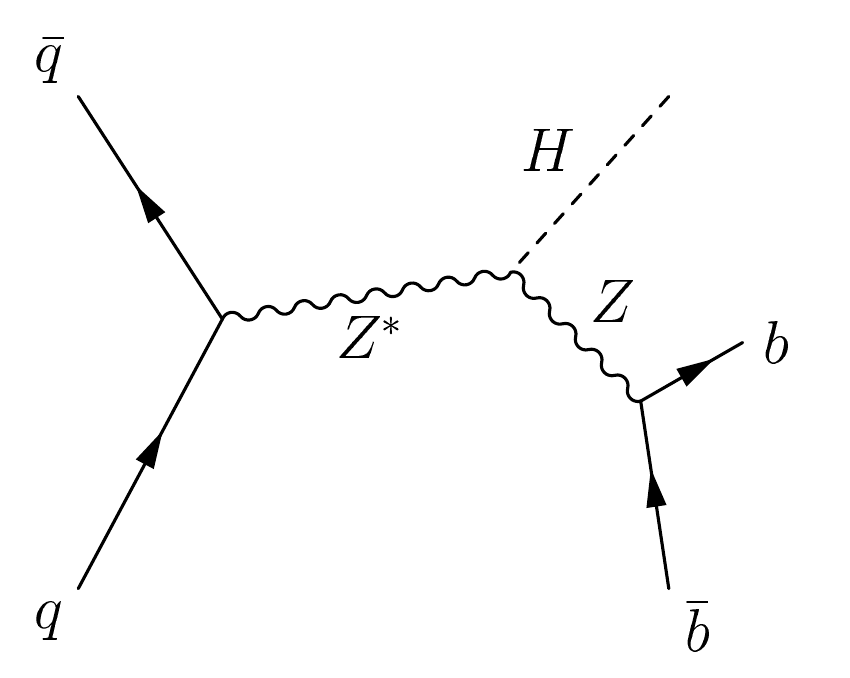}
	\end{center}
	\caption{The Feynman diagrams for Higgs production in the VBF (left), \ZllH\ (center) and \ZbbH\ (right) channels.  The Higgs boson is assumed to decay invisibly.}
	\label{fig:feyn-invisible}
\end{figure*}

Here we report searches for \Hinv\ in the ZH mode, where the Z boson decays to leptons or a \bbbar quark pair, and the first search for \Hinv\ in the vector boson fusion (VBF) production mode, where the Higgs boson is produced in association with two quarks, as shown in Fig.~\ref{fig:feyn-invisible} (left).  Although the VBF signal benefits from a relatively large SM cross section, the final state of two jets plus missing transverse energy (\ETm) suffers from large backgrounds.  However, the backgrounds can be controlled by utilizing the distinct topology of the VBF process, in which the two jets are produced in a forward/backward configuration, with large invariant mass, and are well separated in rapidity.  In addition, hadronic activity in the rapidity gap between the two scattered quarks is reduced, due to the absence of color flow in the VBF process.  The ZH signal, shown in Fig.~\ref{fig:feyn-invisible} (center) and (right), provides a complementary search to the VBF analysis.  Despite a lower SM production cross section, the final state of a Z boson with large \ETm\ provides a clear topology with much lower backgrounds.  We maximize the sensitivity of the search by including decays of the Z boson to leptons and \bbbar quark pairs, which we refer to as \ZllHinv\ and \ZbbHinv\, respectively, where $\ell$ represents either an electron or a muon.  The Higgs boson production modes we consider here rely only on the Higgs boson coupling to the electroweak vector bosons.  New physics that introduces invisible decays of the Higgs boson may also modify these couplings.

In the following sections of this article, we present a brief overview of the Compact Muon Solenoid (CMS) experimental apparatus, physics object reconstruction and datasets in Sections~\ref{sec:apparatus} to~\ref{sec:reco}, followed by a description of the event selection and background estimation for each of the three search channels in Sections~\ref{sec:vbf} to~\ref{sec:zbbh}.  We then present the results of the searches, and their combination, as upper limits on the production cross section times invisible branching fraction in Section~\ref{sec:limits}.  In Section~\ref{sec:dm} we interpret these cross section upper limits in terms of a Higgs-portal model of dark matter interactions, and we summarize our conclusions in Section~\ref{sec:conclusion}.

\section{The CMS apparatus}
\label{sec:apparatus}

The central feature of the CMS apparatus is a superconducting solenoid of 6\unit{m} internal diameter, providing a magnetic field of 3.8\unit{T}. Within the volume of the superconducting solenoid are a silicon pixel and strip tracker, a lead tungstate crystal electromagnetic calorimeter (ECAL), and a brass-scintillator hadron calorimeter, each composed by the barrel and endcap detectors. Muons are measured with detection planes made using three technologies: drift tubes, cathode strip chambers, and resistive-plate chambers, embedded in the steel flux-return yoke outside the solenoid. Extensive forward calorimetry complements the coverage provided by the barrel and endcap detectors. Data are selected online using a two-level trigger system. The first level, consisting of custom made hardware processors, selects events in less than 1\mus, while the high-level trigger processor farm further decreases the event rate from around 100\unit{kHz} to a few hundred Hz before data storage.
The CMS experiment uses a right-handed coordinate system, with the origin at the nominal interaction point, the $x$ axis pointing to the center of the LHC, the $y$ axis pointing up (perpendicular to the LHC plane), and the $z$ axis along the counterclockwise-beam direction. The polar angle $\theta$ is measured from the positive $z$ axis and the azimuthal angle $\phi$ is measured in the $x$-$y$ plane.  The pseudorapidity, $\eta$, is defined as $- \ln [\tan(\theta/2)]$. A more detailed description of the CMS apparatus can be found in Ref.~\cite{Chatrchyan:2008aa}.

\section{Data samples and Monte Carlo simulation}
\label{sec:datasets}

The analyses presented here all use the 8\TeV data sample collected by the CMS Collaboration during 2012, corresponding to an integrated luminosity of 19.5\fbinv in the VBF channel, 19.7\fbinv in the \ZllHinv\ channel, and 18.9\fbinv in the \ZbbHinv\ channel.  The \ZllHinv\ channel also uses the 7\TeV dataset collected during 2011, corresponding to 4.9\fbinv.  The uncertainty assigned to the luminosity measurement is 2.6\% (2.2\%) at $\sqrt{s}=8$ (7)\TeV~\cite{CMS-PAS-LUM-13-001}.  Backgrounds arising from sources other than pp collisions are suppressed using a set of filters that remove events due to anomalous calorimeter signals, beam halo identified in the muon endcaps, inoperable calorimeter cells, and tracking failure.  We further require a well reconstructed vertex within the interaction region; $\abs{z}<24$\cm, $r<2$\cm, where $r=\sqrt{x^{2}+y^2}$.

{\tolerance=1000
The VBF signal is simulated using the \POWHEG 2.0 event generator~\cite{Nason:2004rx,Frixione:2007vw,Alioli:2009je,Hamilton:2009za,Nason:2009ai,Alioli:2010xd,Re:2010bp}, while the \ZllHinv\ and \ZbbHinv\ signals are simulated with \PYTHIA 6.4.26~\cite{Sjostrand:2006za}.  The background processes are simulated using \MADGRAPH 5.1.1~\cite{Alwall:2011uj}, with the exception of some minor backgrounds---specifically, the \VHbb\ background to the \ZbbHinv\ analysis is simulated with \POWHEG 2.0, the diboson backgrounds in the VBF analysis are simulated with \PYTHIA 6.4.26, and the single-top-quark backgrounds in the VBF and \ZllHinv\ analyses use \POWHEG 1.0.  The QCD multijet background is simulated with \PYTHIA 6.4.26.  All samples use the leading-order CTEQ6L1 parton distribution functions (PDFs)~\cite{Pumplin:2002vw}, apart from the \VHbb\ \POWHEG samples, which use the next-to-leading-order (NLO) CTEQ6M PDFs~\cite{Pumplin:2002vw}.  Where yields are estimated directly from MC simulation, the PDF uncertainty is estimated using the PDF4LHC prescription~\cite{Botje:2011sn,Alekhin:2011sk}.  For all Monte Carlo (MC) samples, the detector response is simulated using a detailed description of the CMS detector based on the \GEANTfour package~\cite{Agostinelli:2002hh}.  Minimum bias events are superimposed on the generated events to simulate the effect of multiple pp interactions per bunch crossing (pileup).  Simulated events are weighted such that the distribution of the number of pileup interactions reproduces that observed in data.  The mean number of pileup interactions per bunch crossing was approximately 9 in 2011, and 21 in 2012.  Additional weights are applied to simulated events to ensure trigger efficiency, lepton identification efficiency, and b-tagging efficiency match measurements from data.\par}

The VBF and $\Z\PH$ production cross sections are taken from Refs.~\cite{Dittmaier:2011ti,Dittmaier:2012vm}. The ZH searches are performed in the boosted regime, where the Higgs boson has significant transverse momentum (\pt), and thus, potential differences in the \pt\ spectrum of the \Z\ and Higgs bosons between data and MC generators could introduce systematic effects in the signal acceptance and efficiency estimates.  Two sets of calculations are available that estimate the NLO electroweak corrections~\cite{Ciccolini:2007jr,Ciccolini:2007ec,Denner:2011id} and next-to-next-to-leading order (NNLO) QCD~\cite{Grazzini} corrections to vector boson plus Higgs boson production in the boosted regime.  Both sets of corrections are applied to the signal MC samples. For VH production, the estimated uncertainty arising from the NLO electroweak corrections is 2\%, and from the NNLO QCD corrections is 5\%. In addition, we include NNLO electroweak corrections~\cite{Bierweiler:2013dja} to the $\Z\Z$ and $\PW\Z$ background processes as a function of the \pt\ of the \Z\ boson.

\section{Event reconstruction}
\label{sec:reco}

The reconstructed interaction vertex with the largest value of $\sum_i {\pt}_i^2$, where ${\pt}_i$ is the transverse momentum of the $i$th track associated with the vertex, is selected as the primary event vertex. This vertex is used as the reference vertex for all relevant objects in the event, which are reconstructed with a particle-flow algorithm~\cite{CMS-PAS-PFT-09-001,CMS-PAS-PFT-10-001}. The pileup interactions affect jet momentum reconstruction, missing transverse energy reconstruction, lepton isolation, and \cPqb-tagging efficiencies. To mitigate these effects,  all charged-hadrons that do not originate from the primary interaction are identified by a particle-flow-based algorithm and removed from consideration in the event. In addition, following Ref.~\cite{Cacciari:subtraction}, the average neutral energy density from pileup interactions is evaluated on an event-by-event basis from particle-flow objects and used to compute a correction to the reconstructed jets in the event and to the summed energy in the isolation cones used for leptons.

Muons are reconstructed in the pseudorapidity range $\abs{\eta}< 2.4$.  Two muon reconstruction algorithms are used~\cite{Chatrchyan:2012xi}: one in which tracks in the silicon tracker are matched to signals in the muon detectors, and another in which a global track fit is performed using hits in both the tracker and muon detectors. The muon candidates used in the analysis are required to be successfully reconstructed by both algorithms. The efficiency to reconstruct a muon of $\pt>5$\GeV is larger than 95\%, while the probability to misidentify a hadron as a muon is below 0.1\%. Further identification criteria  are imposed on the muon candidates to reduce the fraction of tracks misidentified as muons. These include the number of measurements in the tracker and in the muon systems, the fit quality of the global muon track and its consistency with the primary vertex.

Electron reconstruction requires the matching of an energy cluster in the ECAL with a track in the silicon tracker~\cite{CMS-PAS-EGM-10-004}. Electron identification relies on a multivariate technique that combines observables sensitive to the amount of bremsstrahlung along the electron trajectory, the geometrical and momentum matching between the electron trajectory and associated clusters, as well as shower-shape observables. Additional requirements are imposed to remove electrons produced by photon conversions. In this analysis, electrons are considered in the pseudorapidity range $\abs{\eta} < 2.5$, excluding the  $1.44 < \abs{\eta}< 1.57$ transition region between the ECAL barrel and endcap, where electron reconstruction is suboptimal.

Jets are reconstructed from particle-flow objects using the anti-\kt clustering algorithm~\cite{Cacciari:2008gp}, with a distance parameter of 0.5, as implemented in the \textsc{fastjet} package~\cite{Cacciari:fastjet1,Cacciari:fastjet2}. Jets are found over the full calorimeter acceptance, $\abs{\eta} < 5$.
Jet energy corrections are applied as a function of the pseudorapidity and transverse momentum of the jet~\cite{Chatrchyan:2011ds}. Jets resulting from pileup interactions are removed using a boosted decision tree (BDT), implemented in the TMVA package~\cite{Hocker:2007ht}, with the following input variables: momentum and spatial distribution of the jet particles, charged- and neutral-particle multiplicities, and consistency of the charged hadrons within the jet with the primary vertex. The missing transverse momentum vector is calculated as the negative of the vectorial sum of the transverse momenta of all particle-flow objects identified in the event, and the magnitude of this vector is referred to as \MET in the rest of this article.

Jets that originate from the hadronization of \cPqb\ quarks are referred to as ``\cPqb\ jets''.  The CSV \cPqb-tagging algorithm~\cite{Chatrchyan:2012jua} is used to identify such jets. The algorithm combines the information about track impact parameters and secondary vertices within jets in a likelihood discriminant to provide separation between \cPqb\ jets and jets originating from light quarks, gluons, or charm quarks. The output of this CSV discriminant has values between zero and one; a jet with a CSV value above a certain threshold is referred to as being ``\cPqb\ tagged''. The efficiency to tag \cPqb\ jets and the rate of misidentification of non-\cPqb\ jets depend on the threshold chosen, and are typically parameterized as a function of the \pt and $\eta$ of the jets. These performance measurements are obtained directly from data in samples that can be enriched in \cPqb\ jets, such as $\ttbar$ and multijet events (where, for example, requiring the presence of a muon in the jets enhances the heavy-flavor content of the events). Several thresholds for the CSV output discriminant are used in this analysis. Depending on the threshold used, the efficiency to tag jets originating from \cPqb\ quarks is in the range 50--75\%, and the probability to incorrectly tag jets originating from \cPqc\ quarks, and light quarks or gluons as \cPqb\ jets are 5--25\%, and 0.15--3.0\%, respectively.

\section{Search for \Hinv\ in vector boson fusion}
\label{sec:vbf}

\subsection{Search strategy}
\label{sec:vbf-strategy}

In the VBF mode, the Higgs boson is produced in association with two final-state quark jets separated by a large rapidity gap, and having high invariant mass.  Loosely following the selection criteria discussed in Ref.~\cite{Eboli:2000ze}, we select final states with two jets and large missing transverse energy and utilize the distinct topology of the VBF jets to discriminate the invisible Higgs boson signal from background.

The dominant backgrounds in this channel result from $\Z (\nu \nu)\text{+jets}$, and $\PW (\ell \nu)\text{+jets}$, where the charged lepton is not identified.  These backgrounds are estimated using control regions with a \Z or \PW\ boson decaying to well identified charged leptons, in association with the same dijet topology used for the signal region.  We then extrapolate from the control regions to the signal region using factors obtained from MC simulation.  The background due to QCD multijet processes, where the \ETm\ arises from mismeasurement, is also estimated from data.  Minor SM backgrounds, arising from $\ttbar$, single-top, diboson, and Drell--Yan$(\ell\ell)\text{+jets}$ processes are estimated from MC simulation.

We use the observed yield in the signal region, together with the estimated background, to perform a single-bin counting experiment.

\subsection{Event selection}
\label{sec:vbf-sel}

We use events collected with a trigger that requires $\ETm>65$\GeV, in association with a pair of jets with $\pt^\mathrm{j1}, \pt^\mathrm{j2} >40\GeV$, in a VBF-like topology.  The jets are required to be in opposite forward/backward halves of the detector, well separated in pseudorapidity ($\etajj = \abs{ \eta_\mathrm{j1} - \eta_\mathrm{j2}} > 3.5$), and with high invariant mass ($\mjj>800$\GeV).  For robustness against pileup, any pair of jets satisfying these criteria is accepted by the trigger.  At the trigger level, the \ETm\ calculation does not include muons, allowing control samples of $\PW(\mu\nu)\text{+jets}$ and $\Z(\mu\mu)\text{+jets}$ events to be taken on the same trigger.  The trigger efficiency is measured in events recorded on a single-muon trigger, as a function of $\pt^\mathrm{j2}$ (since the leading jet, j1, is effectively always above threshold for the regions considered), $\mjj$, and $\ETm$, and the measured efficiency is applied to all MC samples.

The offline selection then proceeds as follows.  We reject backgrounds from \Z and W bosons by vetoing any event with an identified electron~\cite{CMS-PAS-EGM-10-004} or muon~\cite{CMS-PAS-MUO-10-002} with $\pt>10$\GeV.
The VBF tag jet pair is then identified as the leading jet pair.  This pair is required to pass tightened versions of the trigger selection, specifically $\pt^\mathrm{j1},\pt^\mathrm{j2} > 50$\GeV, $\abs{\eta} < 4.7$, $\eta_\mathrm{j1}, \eta_\mathrm{j2} < 0$, $\etajj>4.2$, and $\mjj>1100$\GeV.
The missing-energy requirement is $\ETm > 130$\GeV.  Multijet backgrounds are reduced to a low level by requiring the azimuthal separation between the tag jets to be small, $\phijj < 1.0$~radians, since the background peaks at $\phijj = \pi$~radians while the signal is roughly flat in $\phijj$.  Finally, we apply a central-jet veto (CJV) to any event that has an additional jet with $\pt>30$\GeV and pseudorapidity between those of the two tag jets.

\begin{figure*}[hbtp]
	\begin{center}
		 \includegraphics[width=0.49\textwidth]{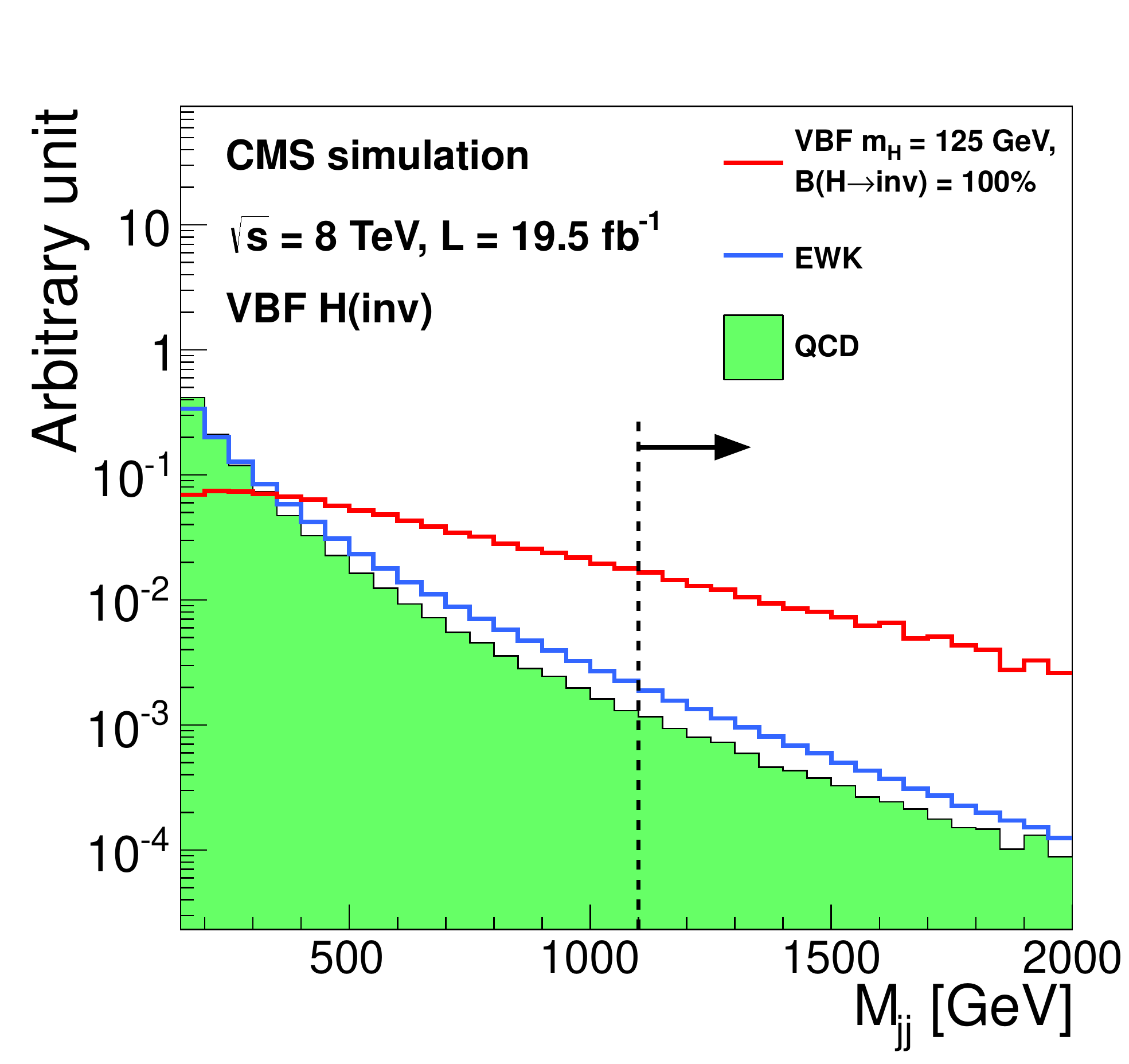}
		 \includegraphics[width=0.49\textwidth]{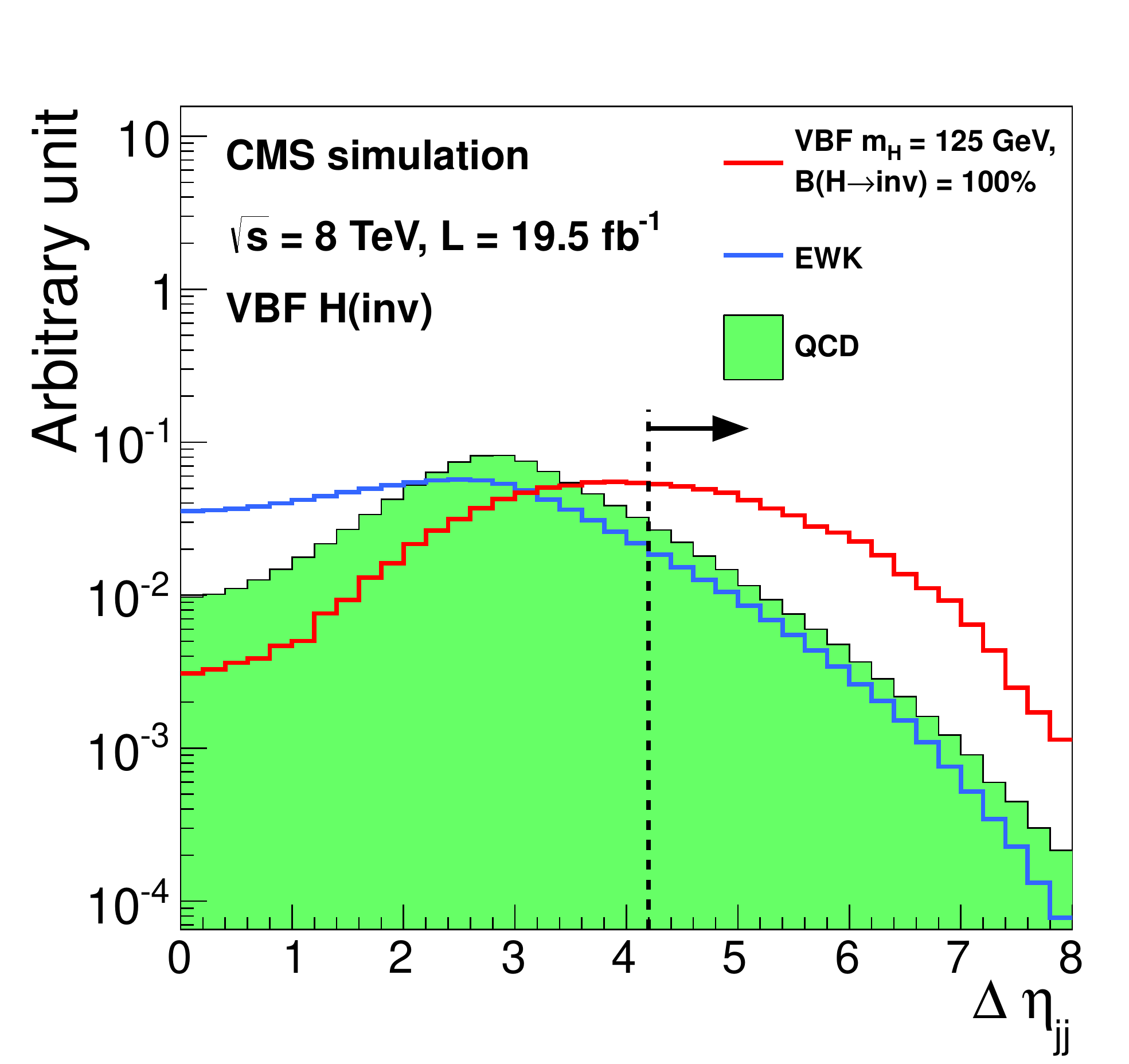}	
		 \includegraphics[width=0.49\textwidth]{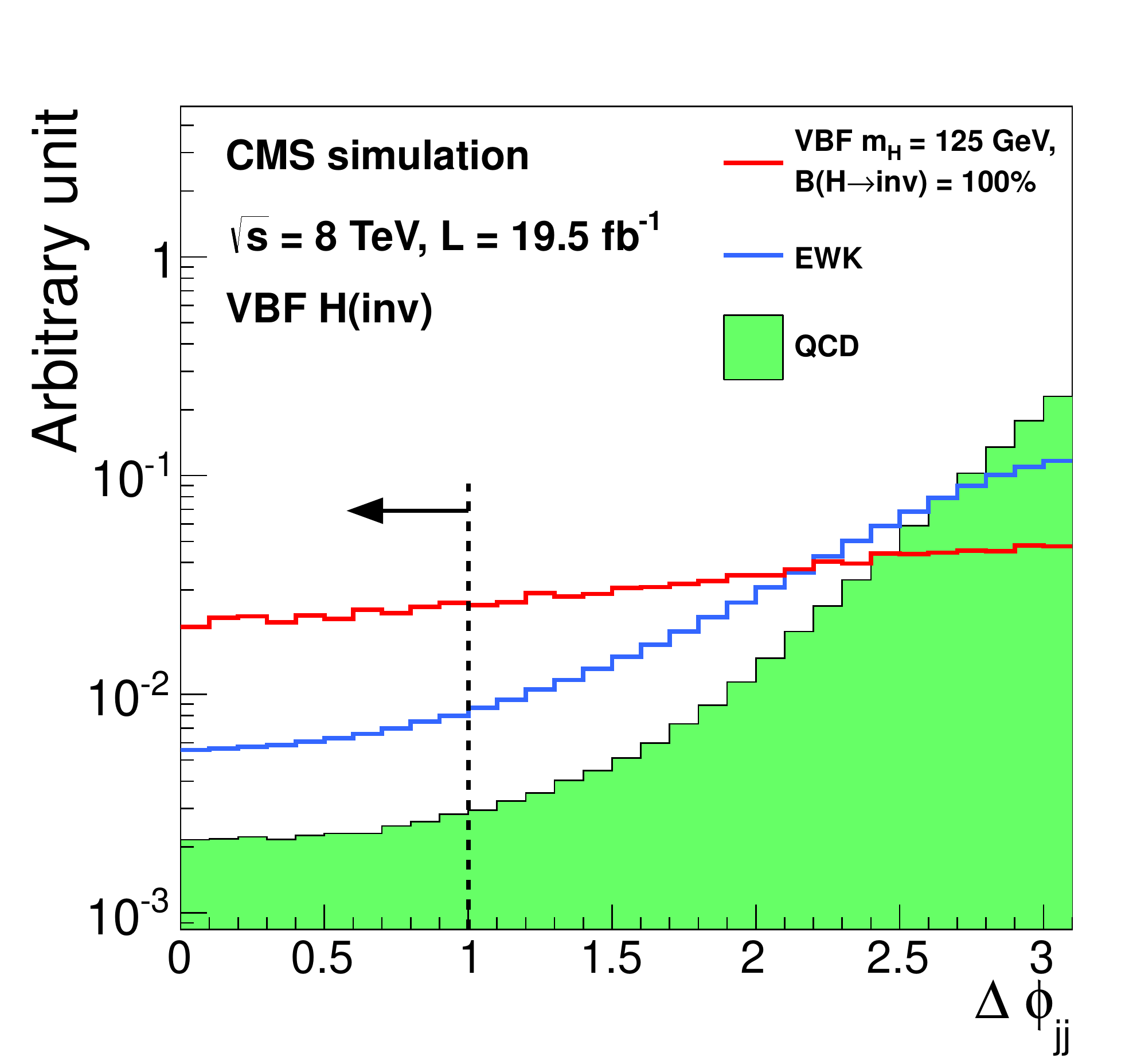}
		 \includegraphics[width=0.49\textwidth]{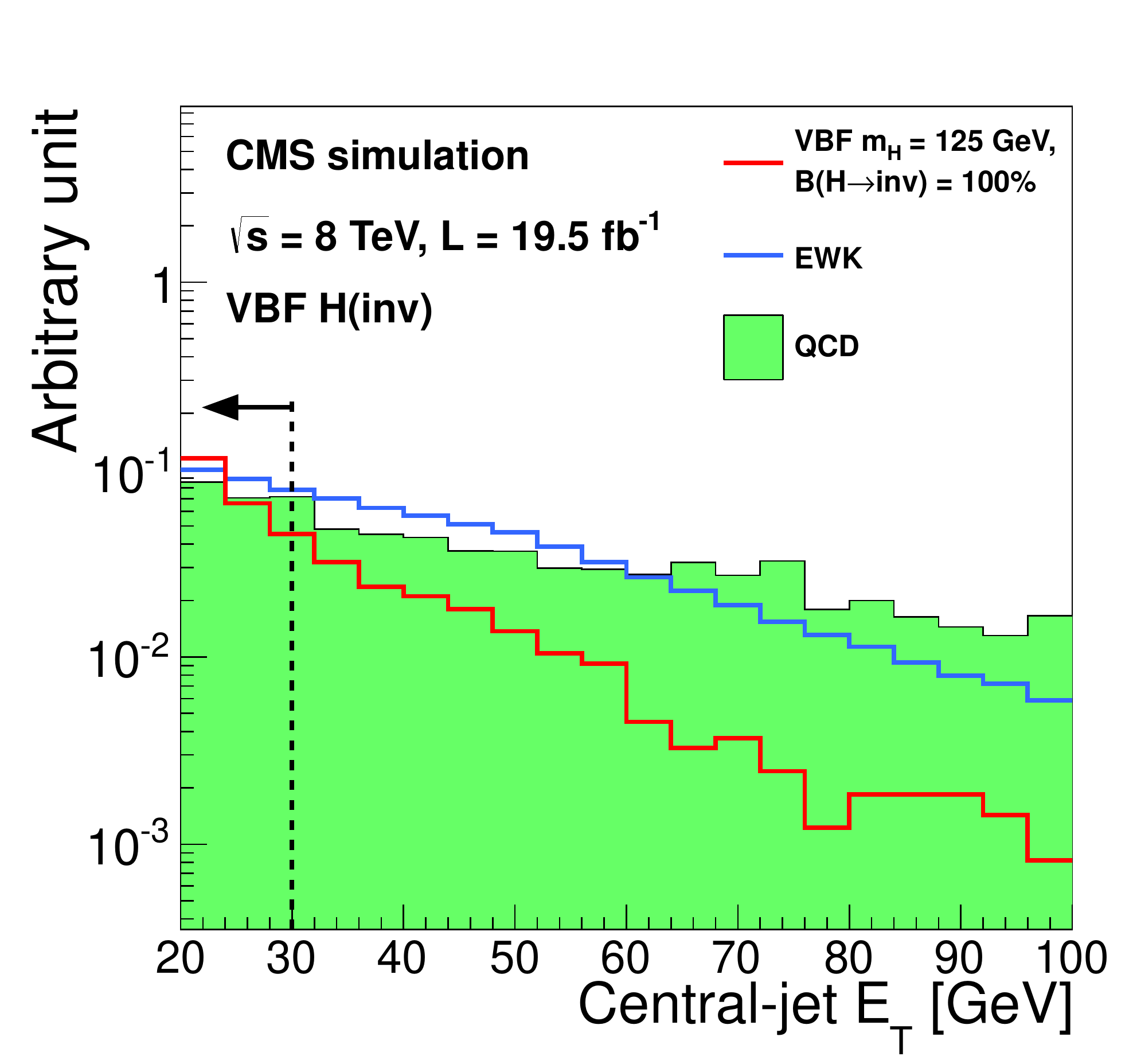}	
		\caption{Distributions of $\mjj$ (top left), $\etajj$ (top right), $\phijj$ (bottom left), and central jet \pt (bottom right) in background and signal MC simulation. The distributions are shown after requiring two jets with $\pt^\mathrm{j1},\pt^\mathrm{j2} > 50$\GeV, $\abs{\eta} < 4.7$, $\eta_\mathrm{j1}, \eta_\mathrm{j2} < 0$, $\mjj>150$\GeV, and $\ETm > 130$\GeV. The arrows correspond to the thresholds applied for the final selection, after optimization.}
		\label{fig:vbfSel}
	\end{center}
\end{figure*}

The lepton and central jet veto thresholds are set to low values at which reconstruction is known to be reliable, while the remaining thresholds are determined by optimizing the selection to give the best signal significance, calculated using a profile likelihood method that incorporates all systematic uncertainties, for a Higgs boson with $\mH=125$\GeV and 100\% invisible branching fraction. The thresholds on jet \pt, $\mjj$, and $\ETm$ are constrained to be above the point where the trigger is 95\% efficient.  This constraint effectively determines the jet \pt and $\ETm$ thresholds, since signal significance only worsens when these thresholds are raised above this point. Distributions of $\mjj$, $\etajj$, $\phijj$, and central jet \pt in background and signal MC simulation are shown in Figure~\ref{fig:vbfSel}, along with the thresholds applied after optimization of the selection.

After all selection requirements, an hypothetical signal equivalent to 125\GeV Higgs boson with $\BRinv=100$\% and produced via the VBF process with SM couplings, is reconstructed with an efficiency of $(6.8 \pm 0.3) \times 10^{-3}$, corresponding to a yield of $210 \pm 29\syst$ events. The requirements on the VBF tag jet $\pt$ and topology, $\mjj$, and $\ETm$ are all correlated and affect the signal efficiency by comparable amounts.  A small signal yield from the gluon-fusion process is also expected, where the VBF requirements may be satisfied by initial-state radiation.  Based on \POWHEG simulation, we estimate this to be $14 \pm 10\syst$ events.

\subsection{Background estimation}
\label{sec:vbf-backgrounds}

The $\Z(\nu \nu)\text{+jets}$ background is estimated from data using observable $\Z (\mu \mu)$ decays. We define a \Z\ control region as for the signal region, with the following changes to the event selection: the lepton veto is replaced with a requirement of an oppositely charged pair of well reconstructed and isolated muons each with $\pt > 20 $\GeV, and invariant mass $60 < M_{\mu\mu}<120$\GeV, a veto is applied on any additional leptons with $\pt>10$\GeV, and the $\ETm$ is recomputed after removing the muons from the \Z\ boson decay. The number of $\Z (\nu \nu)$ events in the signal region is then predicted using:
\begin{equation}
N^\mathrm{s}_{\nu\nu} = (N^\mathrm{c}_{\mu\mu\text{obs}} - N^\mathrm{c}_\text{bkg}) \cdot \frac{\sigma(\Z \to \nu\nu)}{\sigma(\Z/\gamma^{*} \to \mu\mu)} \cdot \frac{\varepsilon^\mathrm{s}_{\Z \mathrm{MC}}}{\varepsilon^\mathrm{c}_{\Z \mathrm{MC}}}.
\end{equation}

\begin{figure}[hbtp]
	\begin{center}
		 \includegraphics[width=0.45\textwidth]{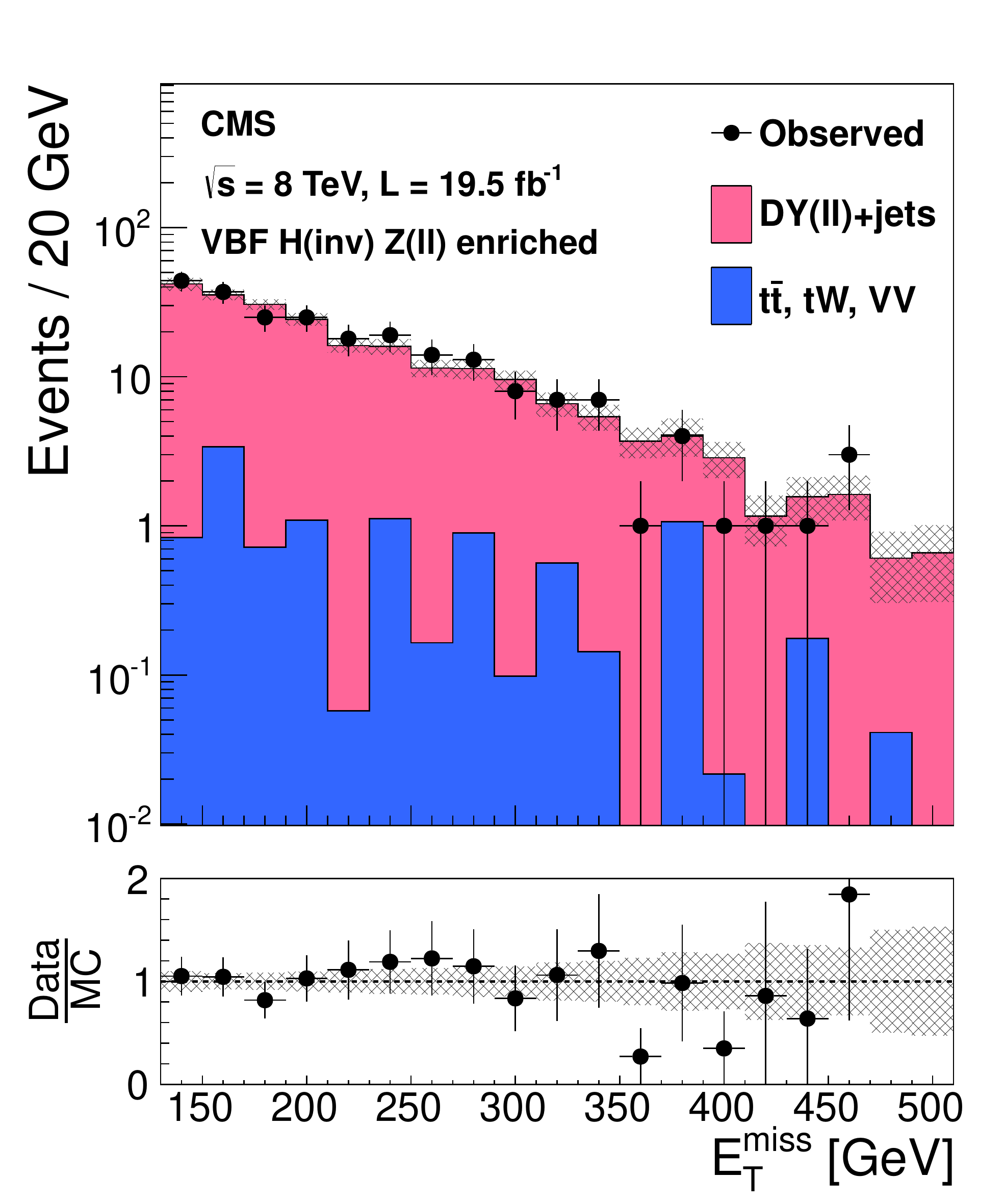}
		 \includegraphics[width=0.45\textwidth]{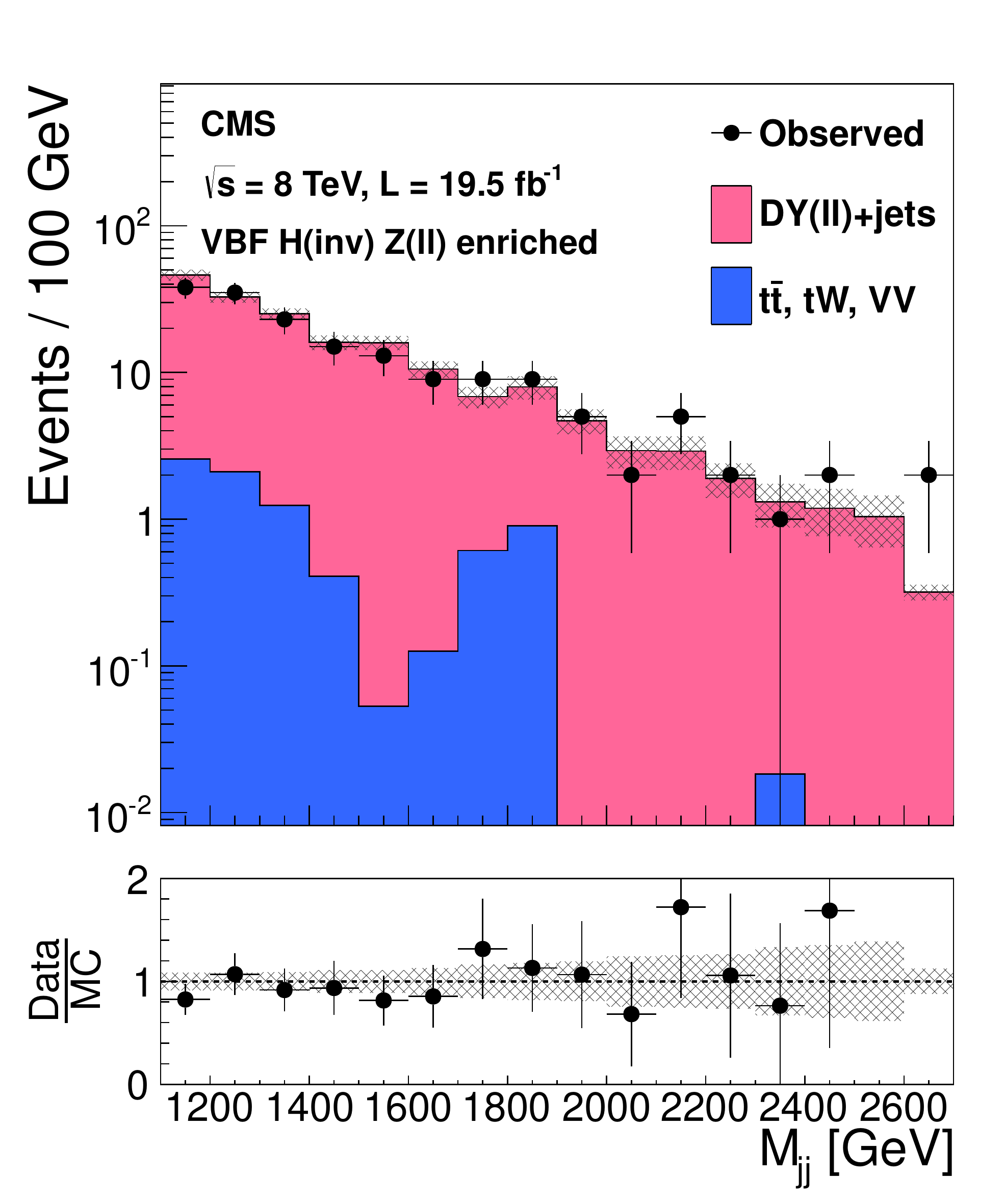}	
		\caption{The $\ETm$ (\cmsLeft) and $\mjj$ (\cmsRight) distributions in the relaxed \Z control region of the VBF search, with no requirements on $\etajj$, $\phijj$, or CJV, and with the $\mjj$ requirement relaxed to 1000\GeV.  The simulated background from different processes is shown cumulatively, and normalized to the data, with its systematic uncertainty shown as a hatched region.  The lower panels show the ratio of data to the simulated background, again with the systematic uncertainty shown as a hatched region.}
		\label{fig:zCtrl}
	\end{center}
\end{figure}

The ratio of cross sections, $\sigma(\Z \to \nu \nu) / \sigma(\Z/\gamma^{*} \to \mu \mu) = 5.651 \pm 0.023\syst$ is calculated with \MCFM~\cite{MCFM} for $m_{\Z/\gamma^{*}} > 50$\GeV, the mass range of the MC sample.  The selection efficiencies in the signal region, $\varepsilon^\mathrm{s}_{\Z \mathrm{MC}} = (1.65 \pm 0.27\syst) \times 10^{-6}$, and the control region, $\varepsilon^\mathrm{c}_{\Z \mathrm{MC}}=(1.11 \pm 0.17\syst) \times 10^{-6}$, are estimated from DY($\ell\ell$)+jets simulation, ignoring the muons when computing the efficiency in the signal region.
The observed yield in the control region is $N^\mathrm{c}_{\mu\mu\text{obs}} = 12$~events.  The background in the control region---estimated from $\ttbar$, diboson and single-top MC samples---is $N^\mathrm{c}_\text{bkg}=0.23 \pm 0.15\syst$~events.  The resulting estimate of the $\Z (\nu \nu)$ background in the signal region is $99 \pm 29\stat \pm 25\syst$ events.  The source of systematic uncertainty in the background estimates will be described in Section~\ref{sec:vbf-syst}.
Figure~\ref{fig:zCtrl} shows the $\ETm$ and dijet invariant mass, $\mjj$, distributions with a relaxed set of criteria for the \Z control region, with $\mjj>1000$\GeV and no requirements on $\etajj$, $\phijj$, or CJV.  In this figure, the simulated background is normalized to the data.  It should be noted that our estimates of the dominant V+jets background are insensitive to the overall normalization of the simulation, which cancels in the ratio.

The $\PW (\Pe \nu)\text{+jets}$ and $\PW (\mu \nu)\text{+jets}$ backgrounds are estimated from single-lepton control samples.
We define $\PW (\mu \nu)$ and $\PW (\Pe \nu)$ control regions in a similar way to the \Z\ boson background.  In the $\PW (\mu \nu)$ region, the lepton veto is replaced with a single $\mu$ requirement and a veto on any additional leptons, and the $\ETm$ is recomputed after removing the muon from the \PW\ boson decay.  The $\PW (\Pe \nu)$ region is defined similarly, with a single electron requirement and additional lepton veto, but here the $\ETm$ is not recomputed, since the electron energy is already included in the $\ETm$ at trigger level.  The number of $\PW (\ell \nu)$ (where ${\ell=\Pe,\mu}$) events in the signal region, $N^\mathrm{s}_{\ell}$ is then estimated using:
\begin{equation}
	\label{eq:w}
N^\mathrm{s}_\ell = (N^\mathrm{c}_{\ell\text{obs}} - N^\mathrm{c}_\text{bkg}) \cdot \frac{N^\mathrm{s}_{\PW \mathrm{MC}}}{N^\mathrm{c}_{\PW \mathrm{MC}}},
\end{equation}
where $N^\mathrm{s}_{\PW \mathrm{MC}}$ and $N^\mathrm{c}_{\PW \mathrm{MC}}$ are the number of events in the signal and control regions in the $\PW (\ell \nu)\text{+jets}$ MC simulation.  The ratio $N^\mathrm{s}_{\PW \mathrm{MC}}/N^\mathrm{c}_{\PW \mathrm{MC}}$ is equal to $0.347 \pm 0.045\syst$ for $\PW (\mu \nu)$ and $1.08 \pm 0.21\syst$ for $\PW (\Pe \nu)$. In the $\PW (\mu \nu)$ control region the observed yield is 223 events, with a background of $30.4 \pm 7.0\syst$ events.  The observed yield in the $\PW (\Pe \nu)$ control region is 65 events, with a background of $7.1 \pm 4.7\syst$ events.  The $\PW (\mu \nu)$ background in the signal region is then estimated to be $66.8 \pm 5.2\stat \pm 15.7\syst$ events, and the $\PW (\Pe \nu)$ background to be $62.7 \pm 8.7\stat \pm 18.1\syst$ events.

The background arising from $\PW (\tau \nu)\text{+jets}$, where the tau lepton decays hadronically ($\tauh$) is estimated using a slightly different method, since a tau lepton veto is not applied in the invisible Higgs boson signal selection.  Hadronically decaying taus are reconstructed using the ``hadron plus strips'' algorithm~\cite{Chatrchyan:2012zz}. This uses charged hadrons and neutral electromagnetic objects (photons) to reconstruct hadronic tau decay modes with one or three charged particles, in the range $\abs{\eta} < 2.3$. A control region is defined, requiring one hadronic tau with $\pt>20$\GeV and $\abs{\eta}<2.3$, no additional leptons, and the remaining signal region selection. However, in the $\PW (\tauh \nu)$ control region, the CJV is not applied in order to increase the yield.  The number of $\PW (\tauh \nu)$ events in the signal region, $N^\mathrm{s}_{\tauh}$, is then estimated from the control region in the same way as the $\PW (\mu \nu)$ and $\PW (\Pe \nu)$ backgrounds. A yield of 32 events is observed in the control region, with the background estimated from the MC simulation to be $15.2 \pm 3.6\syst$ events, giving an estimate of the $\PW (\tauh \nu)$ background in the signal region of $53 \pm 18\stat \pm 18\syst$ events.

In order to cross check the backgrounds from V+jets processes (where V represents either a \PW\ or a \Z boson), which dominate in the signal region, the $\PW (\mu \nu)$ control region and MC simulation is used to compute yields in other control regions. For example, the yield in the $\Z (\mu \mu)$ region is given by:
\begin{equation}
	\label{eq:zfromw}
N^\mathrm{c}_{\mu \mu} = (N^\mathrm{c}_{\mu \text{obs}} - N^\mathrm{c}_\text{bkg}) \cdot \frac{N^\mathrm{c}_{\Z \mathrm{MC}}}{N^\mathrm{c}_{\PW \mathrm{MC}}},
\end{equation}

Similar expressions are used to estimate yields in the $\PW (\Pe \nu)$ and $\PW (\tauh \nu)$ control regions. In all cases, the predictions from data agree with the observed yield within the uncertainty.

The QCD multijet background in the signal region is estimated using the fractions of events passing the $\ETm$ and CJV requirements.  We define regions A, B, C, and D as follows, after the full remaining selection :

\begin{itemize}
	\item{A: fail $\ETm$ selection, fail CJV selection;}
	\item{B: pass $\ETm$ selection, fail CJV selection;}
	\item{C: fail $\ETm$ selection, pass CJV selection;}
	\item{D: pass $\ETm$ selection, pass CJV selection.}
\end{itemize}

We estimate the QCD multijet component in regions A, B, and C from data, after subtracting the electroweak backgrounds using estimations from simulation.  The QCD multijet component in the signal region D can then be estimated using $N_\mathrm{D} = N_\mathrm{ B}N_\mathrm{C} / N_\mathrm{A}$, where $N_{i}$ is the number of events in region $i$.  This method is based on the assumption that the $\ETm$ and the CJV are uncorrelated, which has been checked by comparing the $\ETm$ distribution, below the 130\GeV threshold, in events passing and failing the CJV.  The maximum difference in the $\ETm$ distribution between these two samples is 40\%, which is assigned as a systematic uncertainty of the method.  We predict the QCD background in the signal region to be $30.9 \pm 4.8\stat \pm 23.0\syst$ events.  Furthermore, the method is tested on a high statistics sample with selections equivalent to those in the signal region, but dominated by QCD multijet events by changing the $\phijj$ requirement to $\phijj>2.6$~radians.  In this sample, we observe $2551 \pm 57\stat$ events in the pseudo-signal region after subtraction of backgrounds, which are estimated from MC simulation.  The QCD multijet component is predicted to be $2959 \pm 58\stat$, which is compatible with the observation within the systematic uncertainty.  To give further confidence in this estimate, we perform a cross-check using an ABCD method based on the $\ETm$ and $\phijj$ variables, which gives a prediction consistent with the main method.

The remaining SM backgrounds in the signal region---due to $\ttbar$, single-top, VV and DY($\ell\ell$)+jets---are estimated from MC simulation to be $20.0 ^{+6.0}_{-8.2}\syst$ events.  The total expected background is $332 \pm 36\stat \pm 45\syst$.  The background estimates are summarised in Table~\ref{tab:bgSummary} along with the expected yield for a signal with $\mH=125$\GeV and $\BRinv=100$\%.

\begin{table*}[th!]
	\centering
		\caption{Summary of the estimated number of background and signal events, together with the observed yield, in the VBF search signal region.  The signal yield is given for $\mH=125$\GeV and $\BRinv=100$\%.}
		\label{tab:bgSummary}
		\begin{tabular}{lc}
			\hline \hline
			Process 	 					& Event yields \\
			\hline$\Z (\nu\nu)\text{+jets}$ & $99 \pm 29\stat \pm 25\syst$ \\
			$\PW (\mu\nu)\text{+jets}$ 	 	& $67 \pm 5\stat \pm 16\syst$ 	\\
			$\PW (\Pe \nu)\text{+jets}$  	& $63 \pm 9\stat \pm 18\syst$ 	\\
			$\PW (\tauh \nu)\text{+jets}$ 	& $53 \pm 18\stat \pm 18\syst$ 	\\
			QCD multijet 	 				& $31 \pm 5\stat \pm 23\syst$ 	\\
			Sum (\ttbar, single top quark, $VV$, DY) 	 & $20.0 \pm 8.2\syst$ \\
			\hline
			Total background 				& $332 \pm 36\stat \pm 45\syst$  \\
			VBF H(inv.)						& $210 \pm 29\syst$ \\
			ggF H(inv.)						& $14 \pm 10\syst$ \\
			Observed data					& 390  \\
			\hline
			S/B								& 70\% \\
			\hline \hline
		\end{tabular}
\end{table*}

\subsection{Systematic uncertainty}
\label{sec:vbf-syst}

The V+jets background estimates are affected by large statistical uncertainties, ranging from 5--30\%, due to control samples in data.  The systematic uncertainty in the V+jet background estimates is dominated by the statistical uncertainty in the MC samples used to calculate the control-to-signal region translation factors.  Additional important uncertainties arise due to jet and $\ETm$ energy scale and resolution. These are estimated by varying the scales and resolutions associated with jets and unclustered energy within their uncertainties and recomputing the \ETm, resulting in a 13\% systematic uncertainty in the signal acceptance; 7--15\% in the V+jets background estimates; and 60\% uncertainty in the QCD multijet background estimate. We assign a further 40\% uncertainty to the QCD background estimate, as described in Section~\ref{sec:vbf-backgrounds}.  Although the uncertainty on the QCD background is large, it is a small component of the total background. Small uncertainties in the muon and electron efficiency arise from uncertainties on the scale factors used to correct MC simulation to data, mentioned in Section~\ref{sec:datasets}. For the minor backgrounds estimated from MC, the dominant uncertainties are those associated with the cross sections, which are set according to the corresponding CMS cross section measurements, and the jet/$\ETm$ scale uncertainties.  We consider theoretical uncertainties in the vector boson fusion signal yield resulting from PDF uncertainties and factorization and renormalization scale uncertainties. The uncertainty in the gluon fusion signal yield is dominated by MC modelling of initial-state radiation, amongst other effects, and is estimated to be 60\% by comparing different MC generators. This has a modest overall effect since the gluon fusion yield is small.  These uncertainties are summarized in Table~\ref{tab:syst-qqH}, where they are quoted with respect to the total background or signal yield.  The combined effect of all background uncertainties results in a relative increase of about 65\% in the expected upper limit on the \BRinv.

\begin{table*}[h!t]
\centering
\topcaption{Summary of the uncertainties in the total background and signal yields in the VBF channel. All uncertainties affect the normalization of the yield, and are quoted as the change in the total background or signal estimate, when each systematic effect is varied according to its uncertainties. The signal uncertainties are given for $\mH=125$\GeV and $\BRinv=100$\%.}
\begin{tabular}{lcc}
	\hline \hline
	Source								& Total background & Signal	\\
	\hline
	Control region statistics			& 11\%		& \NA	\\
	MC statistics						& 11\%		& 4\% \\
	Jet/\ETm\ energy scale/resolution	& 7\%		& 13\%	\\
	QCD background estimation			& 4\%		& \NA \\
	Lepton efficiency					& 2\%		& \NA	\\
	Tau ID efficiency					& 1\%		& \NA		 \\
	Luminosity							& 0.2\%		& 2.6\%	\\
	Cross sections						& 0.5--1\%	& \NA \\
	PDFs								& \NA		& 5\% 	 \\
	Factorization/renormalization scale	& \NA		& 4\%    \\
	Gluon fusion signal modelling		& \NA 		& 4\% \\
	\hline
	Total								& 18\%		& 14\% \\
	\hline \hline
\end{tabular}
	\label{tab:syst-qqH}
\end{table*}

\subsection{Results}
\label{sec:vbf-results}

\begin{figure}[hbtp]
	\centering
		 \includegraphics[width=0.49\textwidth]{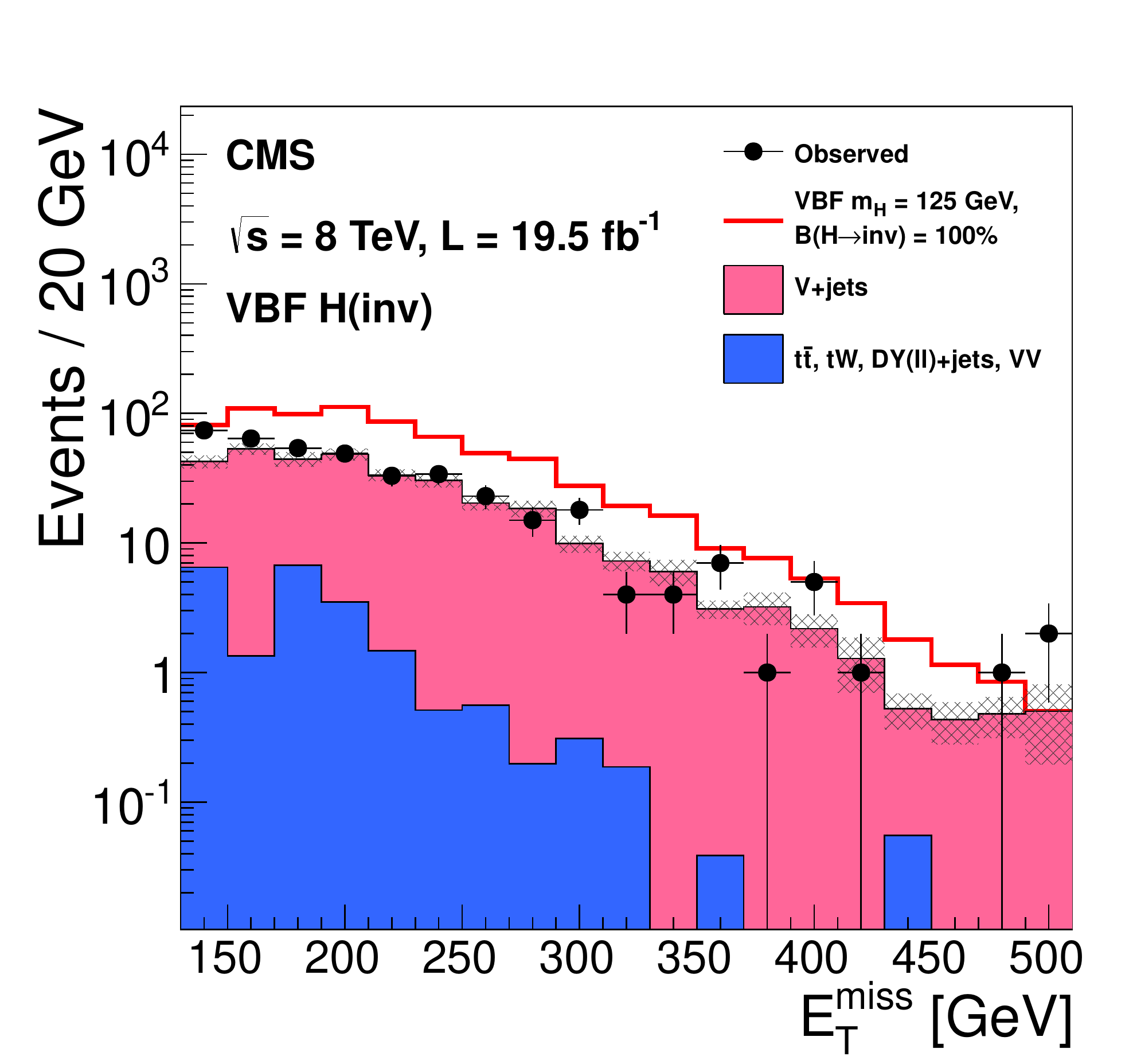}
		 \includegraphics[width=0.49\textwidth]{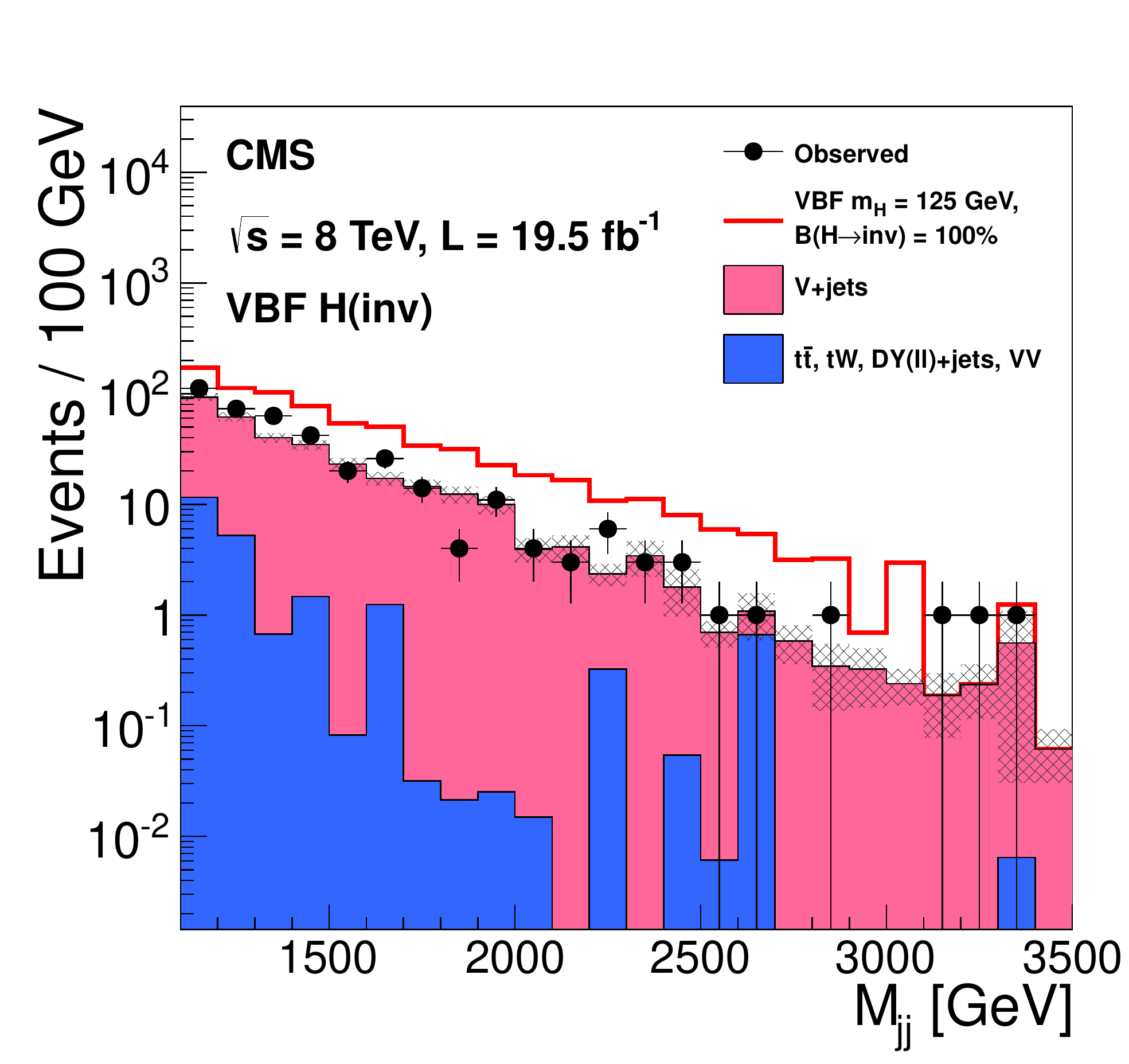}	
		\caption{The $\ETm$ (\cmsLeft) and $\mjj$ (\cmsRight) distributions in data and MC after the full selection in the VBF search signal region.  The simulated background from different processes is normalized to the estimates obtained from control samples in data, and shown cumulatively, with the total systematic uncertainty shown as a hatched region.  Note that the QCD multijet background is not shown due to limited MC statistics, which results in a small apparent discrepancy between data and the backgrounds shown at low values of $\ETm$ and $\mjj$.  The cumulative effect of a signal from a Higgs boson with SM VBF production cross section, $\mH=125$\GeV and \BRinv=100\% is also shown.}
		\label{fig:vbfSignalRegion}
\end{figure}

As shown in Table~\ref{tab:bgSummary}, we observe 390 events the signal region in data, compatible with the background only prediction. Figure~\ref{fig:vbfSignalRegion} shows the $\ETm$ and $\mjj$ distributions in data and simulated backgrounds in the signal region.  The simulated V+jets backgrounds shown in this figure are normalized to the estimates from data given in Table~\ref{tab:bgSummary}.

\section{Search for \texorpdfstring{\ZllHinv}{Z(ll) H(inv)}}

\subsection{Search strategy}
\label{sec:zllh-strategy}

The final state in the \ZllHinv\ channel consists of a pair of high-$\pt$ isolated leptons from the \Z boson decay, high $\ETm$ from the undetectable Higgs boson decay products, and limited jet activity. Since the signal cross section is orders of magnitude lower than those for inclusive DY+jets, $\Wjets$, and $\ttbar$, stringent requirements are needed to isolate the signal. We apply an event selection that is optimized for \mH=125\GeV while still being suitable for the other Higgs boson mass values considered.  After this selection, the dominant backgrounds arise from ZZ and \PW\Z processes, which are modelled using MC simulation. Smaller background contributions, from DY+jets, \ttbar, \PW\PW, and \PW+jets, are modelled using control regions in data.  For each value of the Higgs boson mass, the final background and signal yields used to calculate limits are obtained from a fit to the two-dimensional distribution of the transverse mass, $\mt$, of the dilepton-$\ETm$ system, and the azimuthal separation of the two leptons.

\subsection{Event selection}
\label{sec:zllh-sel}

We use dielectron and dimuon triggers with $\pt>17$\GeV ($\pt>8$\GeV) thresholds for the leading (subleading) lepton, together with single-muon triggers that allow recovery of some residual trigger inefficiencies.  For data taken during periods when the instantaneous luminosity was low enough to allow it, we also use a dimuon trigger with a $\pt>7$\GeV threshold for each muon.

The offline selection starts by requiring two well-identified, isolated leptons of the same flavor and opposite sign ($\Pep\Pem$ or $\Pgmp\Pgmm$), each with $\pt > 20\GeV$.  The invariant mass of the pair must be within $\pm$15\GeV of the $\cPZ$ boson mass.  To reduce the large potential background from DY($\ell\ell$)+jets events, where the $\ETm$ arises from mismeasurement, any event containing two or more jets with $\pt>30\GeV$ is rejected.  The remaining zero- and one-jet samples are treated separately in the analysis because of their significantly different signal-to-background ratios.

The top-quark background is further suppressed by rejecting events containing a bottom-quark decay identified by either the presence of a soft-muon or by the CSV b-tagging algorithm described in Section~\ref{sec:apparatus}. The tagged b jet is required to have $\pt>$20~$\GeV$ and to be reconstructed within the tracker acceptance volume (\ie $\abs{\eta} < 2.5$). The soft-muon is required to have $\pt > 3\GeV$.

To reduce the \WZ\ background in which both bosons decay leptonically, events containing additional electrons or muons with $\pt > 10\GeV$ are rejected.  After all selection requirements, most of the remaining \WZ\ background is from the decay mode $(\PW\to\tau\nu)(\Z\to\ell\ell)$.

The remaining event selection uses three variables:
$\ETm$, $\Delta \phi({\ell\ell,\ETm})$, and $|\ETm-\pt^{\ell\ell}|/\pt^{\ell\ell}$,
where $\pt^{\ell\ell}$ is the transverse momentum
of the dilepton system.
The last two variables effectively suppress reducible background
processes like DY($\ell\ell$)+jets and top-quark production.
We optimized the selection criteria applied to these variables, in order to obtain the best expected exclusion limits at 95\% CL for \mH=125\GeV. For each possible set of
selections, we repeat the full
analysis, including the shape fits described in Section~\ref{sec:zllh-results} below, the estimation of backgrounds
from control data samples,
and the systematic uncertainties.
The final selection criteria obtained after optimization are:
$\ETm > 120\GeV$, $\Delta \phi({\ell\ell,\ETm}) > 2.7$ and
$\abs{\ETm-\pt^{\ell\ell}}/\pt^{\ell\ell} < 0.25$.  The efficiency of the full selection for the \ZllHinv\ signal at $\mH=125$\GeV is 5.6\%, estimated from MC simulation.

\subsection{Background estimation}
\label{sec:zllh-backgrounds}

After the full selection, the dominant backgrounds arise from WZ and ZZ processes, which are modeled using MC simulation.  The pre-fit normalization of these backgrounds is obtained from their respective NLO cross sections computed with \MCFM.

The DY($\ell\ell$)+jets background is modeled from an orthogonal control sample of events
with a single isolated photon produced in association with jets ($\Pgg+\text{jets}$).
This choice has the advantage of providing a large statistics sample,
which resembles $\cPZ$ boson production in all important aspects: production mechanism,
underlying event conditions, pileup scenario, and hadronic recoil~\cite{Chatrchyan:2013oda}.
The kinematic distributions and overall normalization of the $\Pgg+\text{jets}$ events
are matched to $\cPZ(\ell\ell)+\text{jets}$ in data through event weights, determined
as a function of the $\cPZ$ boson $\pt$ measured from data.
This procedure takes into account the dependence of the $\ETm$ on the associated hadronic activity.

Further discrepancies can arise due to differences in the pileup distribution of the $\Pgg+\text{jets}$ sample
due to the fact that photon data was collected with triggers whose prescales varied as a function of photon threshold and data-taking period. These are taken into account by further weighting events in the control sample, according to the distribution of number of reconstructed vertices in the signal sample.
The electroweak backgrounds to the control sample, involving photons and neutrinos, are subtracted using predictions from MC simulation.

\begin{figure*}[hbtp]
	\centering
		 {\includegraphics[width=0.32\textwidth]{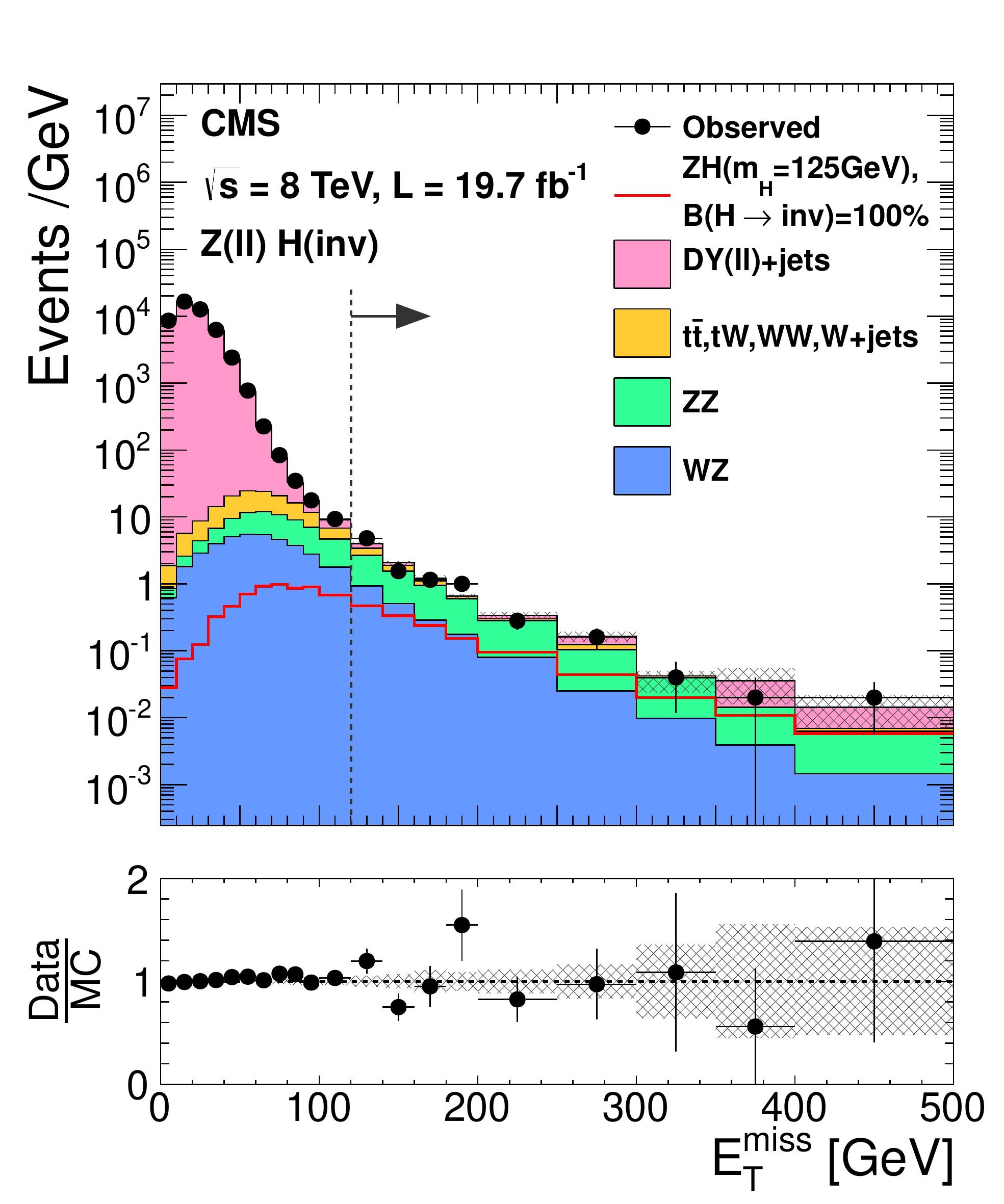}
		 \includegraphics[width=0.32\textwidth]{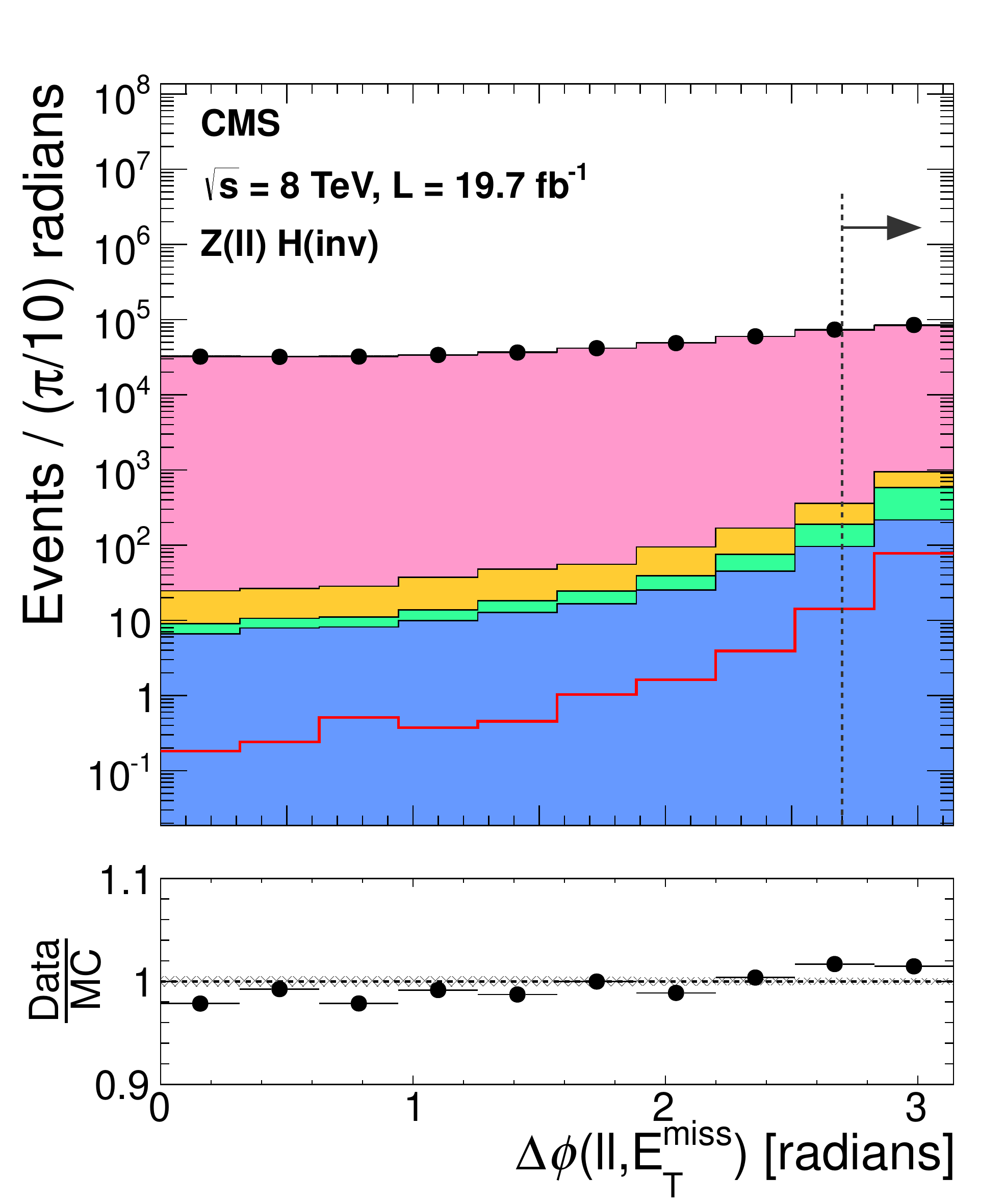}}
		 \includegraphics[width=0.32\textwidth]{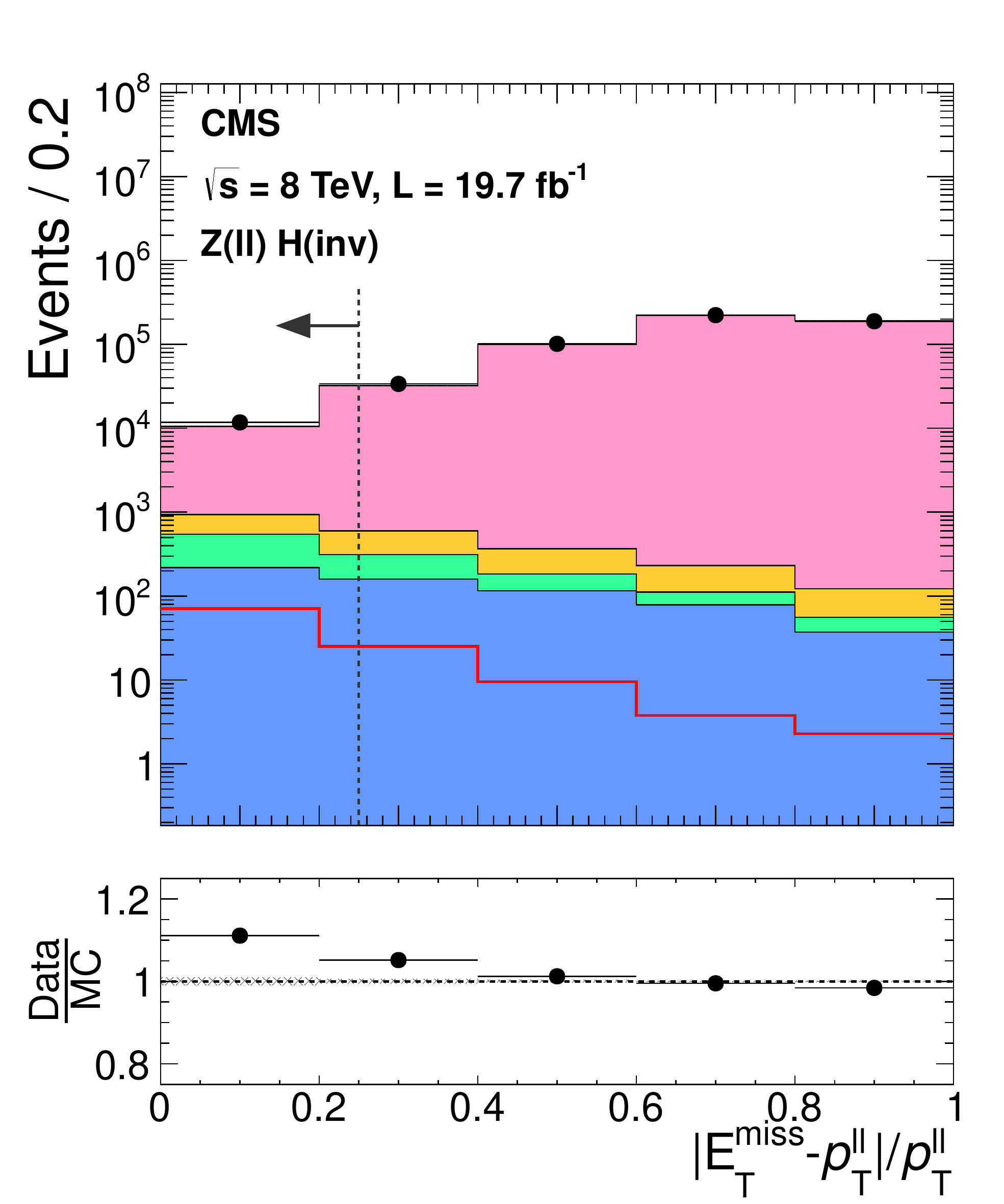}
		\caption{The distributions of $\ETm$ (left), $\Delta \phi({\ell\ell,\ETm})$ (center) , and $|\ETm-\pt^{\ell\ell}|/\pt^{\ell\ell}$ (right) in data compared to the estimated background from simulation (\PW\Z and \ZZ) or data (all other channels), before the optimization of the selection. The expected distributions from different background processes are displayed cumulatively, while a signal corresponding to $\mH=125$\GeV and \BRinv=100\% is superimposed separately. The arrows correspond to the cuts applied for the final selection as described at the end of Section~\ref{sec:zllh-sel}. The statistical uncertainty in the background estimate is shown as a hatched region. The plots show the electron and muon channels combined. The lower panels show the ratio of data to the simulated background, again with the statistical uncertainty in the background shown as a hatched region.}
		\label{fig:zgamma_met_data}
\end{figure*}

This procedure yields an accurate model of the $\ETm$ distribution in DY($\ell\ell$)+jets events,
as shown in Fig.~\ref{fig:zgamma_met_data} (left), which compares the
$\ETm$ distribution of the weighted $\Pgg\text{+jets}$ events, summed with
other backgrounds, to the $\ETm$
distribution of the dilepton events in data.
Figure~\ref{fig:zgamma_met_data} also compares
the distributions of (center) $\Delta \phi(\ell\ell,\ETm)$  and (right) $|\ETm-\pt^{\ell\ell}|/\pt^{\ell\ell}$ obtained from this background model to the same distributions in the dilepton sample.
The difference  between data and background predictions is less than 10\% in these distributions, which is negligible compared to the estimated systematic uncertainties after the final selection.
The uncertainties in the electroweak background to the photon control sample yield a 100\% uncertainty in the normalization of the residual DY($\ell\ell$)+jets background. However, since the Drell--Yan background after the full selection is very small, the large uncertainty has negligible impact on the final results.

The remaining background processes do not involve \Z boson production, and are referred to as non-resonant backgrounds. Such backgrounds arise mainly from leptonic \PW\ boson decays in $\ttbar$, $\cPqt\PW$ decays and $\PW\PW$ events. Also included in the estimate of non-resonant backgrounds are small contributions from single-top-quark events produced from $s$- and $t$-channel processes, $\PW\text{+jets}$ production, and $\Z\to \Pgt\Pgt$ events in which $\Pgt$ leptons produce electrons/muons and \ETm.

We estimate these backgrounds using a control sample in data, consisting of events with opposite-charge different-flavor dilepton pairs ($\Pe^{\pm}\Pgm^{\mp}$) that otherwise pass the full selection.  The backgrounds in the $\Pep\Pem$ and $\Pgmp\Pgmm$ final states are then estimated by applying scale factors ($\alpha_{\Pe\Pe}$, $\alpha_{\Pgm\Pgm}$) to the number of events in the control sample, $N_{\Pe\Pgm}$:
\begin{equation}
N_{\Pe\Pe} = \alpha_{\Pe\Pe} \times N_{\Pe\Pgm}, \qquad
N_{\Pgm\Pgm} = \alpha_{\Pgm\Pgm} \times N_{\Pe\Pgm}.
\end{equation}

We compute the two factors $\alpha_{\Pe\Pe}$ and $\alpha_{\Pgm\Pgm}$  in the sidebands (SB) of the $\cPZ$ peak
($40< \mll <70$\GeV and $110< \mll < 200$\GeV) by using the following relations:
\begin{equation}\alpha_{\Pe\Pe} = \frac{N_{\Pe\Pe}^\mathrm{SB}}{N_{\Pe\Pgm}^\mathrm{SB}}, \qquad
\alpha_{\Pgm\Pgm} = \frac{N_{\Pgm\Pgm}^\mathrm{SB}}{N_{\Pe\Pgm}^\mathrm{SB}},
\end{equation}

where $N_{\Pe\Pe}^\mathrm{SB}$, $N_{\Pgm\Pgm}^\mathrm{SB}$, and $N_{\Pe\Pgm}^\mathrm{SB}$ are the number of events
in the $\cPZ$ sidebands counted in a top-quark-enriched sample of $\Pep\Pem$, $\Pgmp\Pgmm$, and $\Pe^{\pm}\Pgm^{\mp}$ final states, respectively. The requirements for this sample are $\ETm>65$\GeV, $\pt^{\ell\ell} >50$\GeV, $0.4 <\ETm/\pt^{\ell\ell} < 1.8$, and a \cPqb-tagged jet. The kinematic requirements are looser than in the signal region, in order to reduce the statistical uncertainties in the scale factors.
The measured values of these factors with the corresponding statistical uncertainties
are $\alpha_{\Pe\Pe}^{7\TeV} = 0.42 \pm 0.04$, $\alpha_{\Pgm\Pgm}^{7\TeV}= 0.64 \pm 0.06$
and $\alpha_{\Pe\Pe}^{8\TeV} = 0.43 \pm 0.02$, $\alpha_{\Pgm\Pgm}^{8\TeV}= 0.69 \pm 0.03$.
The validity of the procedure for computing the scale factor is checked by closure tests
on simulated samples. This method accounts for possible differences in probability for electrons and muons to pass the trigger and selection requirements. We also cross-check the methods by calculating $\alpha_{\Pe\Pe}$
and $\alpha_{\Pgm\Pgm}$ from the $\cPZ$ peak region as follows:
\begin{equation}
\alpha_{\Pe\Pe} = \frac{1}{2}\sqrt{\frac{N_{\Pe\Pe}^\text{peak}}{N_{\Pgm\Pgm}^\text{peak}}},
\qquad
\alpha_{\Pgm\Pgm} = \frac{1}{2}\sqrt{\frac{N_{\Pgm\Pgm}^\text{peak}}{N_{\Pe\Pe}^\text{peak}}}, \end{equation}

where $N_{\Pe\Pe}^\text{peak}$, $N_{\Pgm\Pgm}^\text{peak}$, are the number of dielectron and dimuon events in a $\cPZ$ control sample. This method takes advantage of the equality between the production rates for $\Z\to\Pe\Pe$ and $\Z\to\Pgm\Pgm$ and equates the ratio of observed dilepton counts to the square of the ratio of efficiencies. From the comparison of methods and the closure tests, we derive an uncertainty of 25\% on the normalization of the non-resonant background in addition to the contribution from the statistical uncertainties on the control samples.  The background in the signal region, estimated using the methods described above, are shown in Table~\ref{tab:zhinvsel}, along with the expected yield for a signal with $\mH=125$\GeV and 100\% invisible branching fraction.

\begin{table*}[h!t]
 \centering
 \topcaption{Observed yields, background estimates and signal predictions at $\sqrt{s}=7$\TeV and 8\TeV in the \ZllHinv\ channel. The signal yields are given for $\mH=125$\GeV and $\BRinv=100$\%.}
\label{tab:zhinvsel}
 \begin{tabular}{ccccc}
\hline \hline
Process & \multicolumn{2}{c}{$\sqrt{s} = 7\TeV$} & \multicolumn{2}{c}{$\sqrt{s} = 8\TeV$} \\
        & $\Pe\Pe$ & $\Pgm\Pgm$ & $\Pe\Pe$ & $\Pgm\Pgm$ \\
\hline \hline
& \multicolumn{4}{c}{0 jet selection} \\
\hline
\hline
$\dyll$                   &  0.1 $\pm$ 0.1     &     0.2 $\pm$ 0.2      	 &    0.2 $\pm$ 0.3    &    0.9 $\pm$ 1.4 \\
$\WZ\to 3\ell\nu$  &  1.7 $\pm$ 0.2     &     2.0 $\pm$ 0.3      	 &   10.4 $\pm$ 1.6    &   14.1 $\pm$ 1.7 \\
$\ZZ\to 2\ell2\nu$ &  5.8 $\pm$ 0.7     &     7.8 $\pm$ 0.9      	 &   26.4 $\pm$ 3.0    &   35.9 $\pm$ 3.6 \\
$\ttbar$, $\Wt$, $\PW\PW$ \& $\PW\text{+jets}$           &  1.1 $\pm$ 6.4     &     1.0 $\pm$ 3.1      	 &    0.4 $\pm$ 1.5    &    0.7 $\pm$ 2.1 \\
\hline
Total backgrounds                &  8.7 $\pm$ 6.5     &    11.0 $\pm$ 3.3      	 &   37.4 $\pm$ 3.7    &   51.6 $\pm$ 4.8 \\
$\Z\PH(125)$                 &  2.3 $\pm$ 0.2     &     3.1 $\pm$ 0.3      	 &   10.3 $\pm$ 1.2    &   14.7 $\pm$ 1.5 \\
Observed data                      &  9                 &     10                 	 &   36                &    46            \\
\hline
S/B                     &  26\%                 & 28\%                       &   28\%                 & 24\% \\
\hline \hline
& \multicolumn{4}{c}{1 jet selection} \\
\hline
\hline
$\dyll$                   &  0.2 $\pm$ 0.2     &     0.0 $\pm$ $^{1.3}_{0.0}$ 	&    2.0 $\pm$ 3.8    &    3.0 $\pm$ 5.6 \\
$\WZ\to 3\ell\nu$  &  0.8 $\pm$ 0.1     &     0.9 $\pm$  0.2     	 &    3.3 $\pm$ 0.4    &    3.8 $\pm$ 0.5 \\
$\ZZ\to 2\ell2\nu$ &  1.1 $\pm$ 0.2     &     1.4 $\pm$ 0.2      	 &    4.8 $\pm$ 0.5    &    6.3 $\pm$ 0.7 \\
$\ttbar$, $\Wt$, $\PW\PW$ \& $\PW\text{+jets}$           &  0.5 $\pm$ 0.6     &     0.5 $\pm$ 0.8      	 &    0.4 $\pm$ 1.7    &    0.7 $\pm$ 1.3 \\
\hline
Total backgrounds               &  2.6 $\pm$ 0.7     &     2.8 $\pm$  0.9     	 &   10.6 $\pm$ 4.2    &   13.8 $\pm$ 5.8 \\
$\Z\PH(125)$                 &  0.4 $\pm$ 0.1     &     0.5 $\pm$ 0.1      	 &    1.6 $\pm$ 0.2    &    2.5 $\pm$ 0.3 \\
Observed data                      &  1                 &     4                  	 &    11               &    17            \\
\hline
S/B                   & 15\%                  & 18\%                       & 15\%                    & 18\% \\
\hline \hline
 \end{tabular}
\end{table*}

\subsection{Systematic uncertainty}
\label{sec:zllh-syst}

Table \ref{tab:syst-ZllH} lists the systematic uncertainties affecting this search.  The most important uncertainties are those associated with theory, affecting both the signal acceptance and the dominant $\PW\Z$ and $\Z\Z$ backgrounds. The uncertainties arising from missing higher-order QCD
corrections are estimated by scaling the renormalization and factorization scales up
and down by a factor of two, while those associated with PDFs are estimated using the PDF4LHC prescription~\cite{Botje:2011sn,Alekhin:2011sk}.

The uncertainties related to jet and $\ETm$ energy scale and resolution, lepton \pt scale, and reconstruction efficiency affect the signal and all backgrounds, and are estimated as for the search in the VBF mode (see Section~\ref{sec:vbf-syst}).

Uncertainties of approximately $100$\%, which are derived from the data by comparing different estimation methods and conducting closure tests, are assigned to the non-resonant backgrounds.  Due to the small size of the control samples, the relative uncertainties are large, but absolute contribution of these backgrounds is small.

The combined signal efficiency uncertainty is estimated to be $\sim$12\%, and the total uncertainty in the
background estimations is about $\sim$15\%, dominated by the theoretical uncertainties mentioned above. The combined effect of all systematic uncertainties results in a relative increase of about 35\% in the expected upper limit on the \BRinv.

\begin{table*}[h!t]
\centering
\topcaption{Summary of systematic uncertainties in the \ZllHinv\ channel. The numbers indicate the change in the total background estimate or in the total signal acceptance when each systematic effect is varied according to its uncertainties. Those uncertainties designated as ``Norm." only affect the normalization of the contributions, while those designated ``Shape" also affect the shapes of  the $\mt$ and/or $\Delta\phi(\ell\ell)$ distributions. In the case of shape variations, the numbers indicate the range of changes across the bins of the distributions. Signal uncertainties are quoted for $\mH=125$\GeV and $\BRinv=100\%$.}
\begin{tabular}{llccc}
  \hline \hline
  Type & Source & Background & Signal \\
 & & uncertainty(\%) & uncertainty(\%) \\
  \hline
\multirow{5}{*}{Norm.}
&  PDFs                                      & 5.0 & 5.7    \\
&  Factorization/renormalization scale  & 6.4 & 7.0     \\
&  Luminosity                               & 2.3 & 2.2--2.6 \\
&  Lepton trigger, reconstruction, isolation         & 2.7 & 3.0     \\
&  Drell--Yan normalization                    & 4.8 & \NA  \\
&  $\ttbar$, $\Wt$, $\PW\PW$ \& $\PW\text{+jets}$ normalization        & 1.0 & \NA \\
	\hline
\multirow{5}{*}{Shape}
&   MC statistics ($\Z\PH$, $\Z\Z$, $\PW\Z$) & 1.8--3.8 & 3.0--4.0    \\
&   Control region statistics (DY($\ell\ell$)+jets)       & 0.6--1.2 & \NA  \\
&   Control region statistics ($\ttbar$, $\Wt$, $\PW\PW$ \& $\PW\text{+jets}$)    & 2.0-3.8 & \NA\\
&   Pile up                              & 0.2 & 0.3 \\
&   b-tagging efficiency		   		& 0.2 & 0.2   \\
&   Lepton momentum scale                & 0.9 & 1.0      \\
&   Jet energy scale/resolution          & 2.4--3.1 & 2.6--3.2    \\
&   \ETm\ scale 					& 1.7--2.9 & 1.4--2.3    \\
	\hline
&	Total							&	11-12		&	11		\\
  \hline \hline
\end{tabular}
\label{tab:syst-ZllH}
\end{table*}

\subsection{Results}
\label{sec:zllh-results}

As shown in Table~\ref{tab:zhinvsel}, the total number of observed events is 134 with an estimated background of about 138 events, while the expected signal yield is 35 events.  The final limits on a signal are determined using a profile likelihood fit to the normalizations and the shapes of selected distributions in the signal region.  For the 8\TeV data, we use the two-dimensional distribution of the azimuthal dilepton separation ($\delphill$) and the $\mt$ of the dilepton-$\ETm$ system. For the 7\TeV data, due to lower number of events in the control samples, we use a one-dimensional fit to $\mt$ alone.  The expected ratio of signal to background increases at high values of $\mt$ and low values of $\delphill$, giving the shape analysis greater sensitivity than a limit obtained from event counts alone.  The transverse mass $\mt$ is given by the formula
\begin{equation}
\mt = \sqrt{2 \pt^{\ell\ell} \ETm \left[ 1-\cos \Delta \phi \left({\ell\ell,\ETm}\right) \right]}. \end{equation}

This definition of $\mt$, which treats both the lepton pair and the recoiling
undetected system as massless, is found to yield the best separation between the
signal and the backgrounds from $\PW\PW$, $\PW\cPZ$, and $\cPZ\cPZ$.

\begin{figure*}[htbp]
	\centering
		 \includegraphics[width=0.32\textwidth]{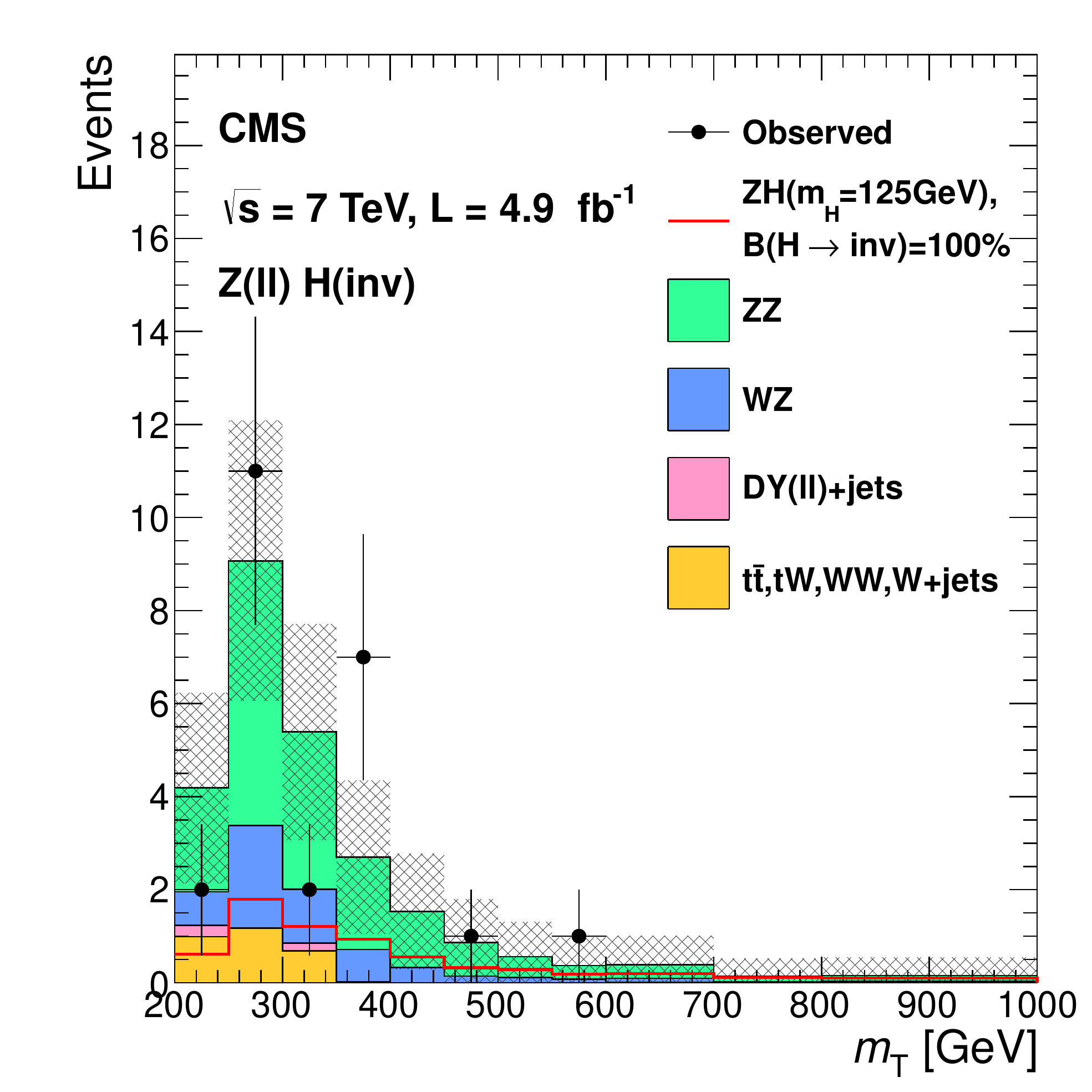}
		 \includegraphics[width=0.32\textwidth]{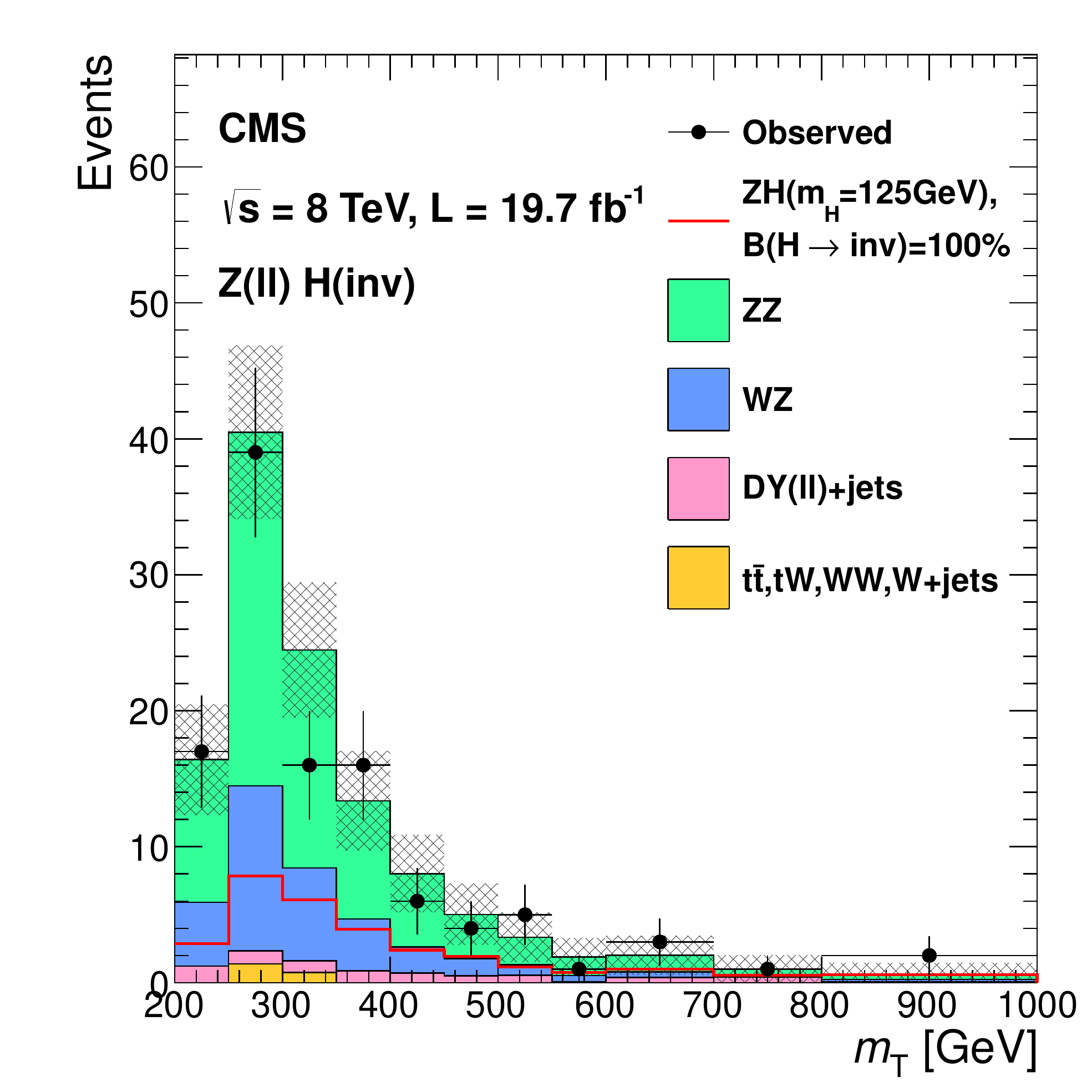}
		 \includegraphics[width=0.32\textwidth]{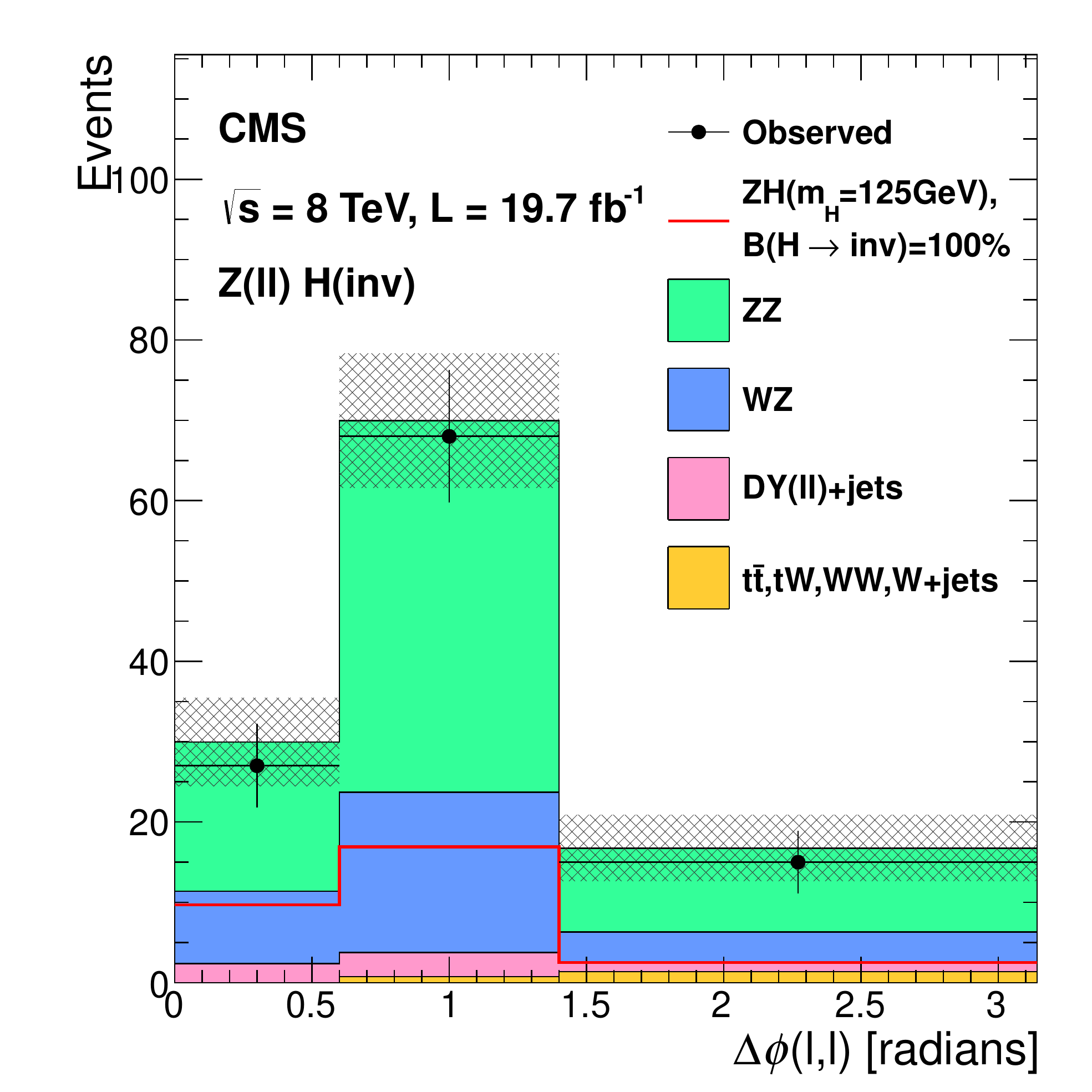}
    	\caption{Distributions used for setting limits in the \ZllHinv\ analysis.
	The expected distributions from different background processes are
	displayed cumulatively, while a signal corresponding to $\mH=125$\GeV and \BRinv=100\% is
	superimposed separately. The total statistical and systematic uncertainty in the
	total background is shown as a hatched region. The limits for 7\TeV use the shape of the $\mt$ distribution (left) while the limits for 8\TeV use both the $\mt$ (center) and $\delphill$ (right) shapes.
	The distributions are shown with  electron and muon channels and 0- and 1-jet
channels combined.}
    	\label{fig:ZllH-lim}
\end{figure*}

The two center-of-mass energies (7 and 8\TeV), two lepton flavors (\Pe\ and \Pgm), and two jet multiplicities (0 and 1), define eight disjoint samples that are treated separately in the likelihood calculation.  The shapes and normalizations of the signal and of each background component are allowed to vary within their uncertainties, and correlations in the sources of systematic uncertainty are taken into account.  The $\mt$ distribution in the 7\TeV data, and the $\delphill$ distribution in the 8\TeV data, in the signal region are shown in Fig.~\ref{fig:ZllH-lim} for illustration.  As can be seen, the observed data are consistent with the predicted backgrounds.

\section{Search for \texorpdfstring{\ZbbHinv}{Z(b b-bar) H(inv)}}
\label{sec:zbbh}

\subsection{Search strategy}
\label{sec:zbbh-strategy}

The \ZbbHinv\ search closely follows the strategy of the CMS search for SM \ZnnH~\cite{Chatrchyan:2013zna}, sharing the same $\ETm+\bbbar$ final state, though the \bbbar resonances have different masses.  The event selection requires large $\ETm$, equivalent to the boost of the Higgs boson~\cite{PhysRevLett.100.242001}, and a jet pair consistent with a \Ztobb\ decay.
The signal yield after the final selection is estimated using a BDT trained on simulated background and signal MC samples, by fitting BDT output for background and signal to that obtained from data.

The backgrounds in this channel arise from production of W and \Z bosons in association with jets (V+jets), \ttbar, single-top-quark, diboson (VV), and QCD multijet production. The SM Higgs process, \ZnnH, has a negligible effect on this search, due to the different mass of the \bbbar resonance and good di-jet mass resolution, which is about 10\%.  The \ZnnH\ process is therefore treated as an independent background process.

Since the VV production cross section is only a small factor larger than that of standard model VH, and given the nearly identical final state for VZ with $\Z(\bbbar)$, the VV process has been used as a benchmark to validate the search strategy used here~\cite{Chatrchyan:2013zna}.

\subsection{Trigger}
\label{sec:zbbh-trigger}

A suite of four \ETm\ triggers is used for this search, due to the challenge of maintaining acceptance as the instantaneous luminosity increases.  A trigger with $\ETm >150$\GeV is used for the full 8\TeV data set. To increase acceptance at lower $\ETm$, we also use triggers requiring jets in addition to $\ETm$.  For the early data-taking period, a trigger requiring $\ETm>80$\GeV together with two jets with $\abs{\eta}<2.5$ and $\pt>30$\GeV was used.  However, as the average instantaneous luminosity reached $3\times 10^{33}\percms$, this was replaced with a trigger requiring $\ETm>100$\GeV, two jets with individual \pt\ above 60 and 25\GeV respectively, the vector sum of the two jet \pt to be above 100\GeV, and finally a veto on any jet with $\pt>40$\GeV and closer than 0.5 radians in $\phi$ to the $\ETm$ direction. Finally, a trigger was used that requires $\ETm>80$\GeV, together with two jets having $\abs{\eta}<2.5$ and $\pt>20$\GeV or $\pt>30$\GeV, depending on the luminosity conditions, and at least one of the jets tagged by the online CSV b-tagging algorithm~\cite{Chatrchyan:2012jua}.

For \ZbbHinv\ events with $\ETm >170$\GeV, the combined trigger efficiency is near 100\% with respect to the offline event reconstruction and selection, described in the next section. For events with \ETm\ between 130 and 170\GeV (100 and 130\GeV) the corresponding efficiency is about 98\% (85\%).

\subsection{Event selection}
\label{sec:zbbh-sel}

{\tolerance=600
The event selection in this channel is designed to enhance heavy-flavor production and a Higgs boson with high Lorentz boost, with reasonable kinematic thresholds consistent with the trigger selection, and to provide sufficient statistics to perform the BDT training properly.  The event selection is summarized in Table~\ref{tab:zbbh-sel}. Backgrounds to the signal are substantially reduced by a large \ETm requirement. In this regime, where the Higgs boson has substantial boost, the Z and Higgs bosons are separated by a large azimuthal opening angle, we therefore require $\dphiZH>2.0$~radians. We define ``low'', ``intermediate'', and ``high'' \ETm\ regions to have $100<\ETm<130$\GeV, $130<\ETm<170$\GeV, and $\ETm>170$\GeV, respectively.
\par}

The QCD multijet background is reduced to negligible levels by imposing three requirements which ensure that the \ETm\ does not originate from mismeasured jets. First, we cut on the azimuthal separation, $\dphiMJ$, between the \ETm\ direction and the closest jet with $\abs{\eta}<2.5$ and \pt$>25$\GeV.  For the high-\ETm\ region we require $\dphiMJ>0.5$ radians, while for the intermediate- and low-\ptV\ regions this requirement is increased to $\dphiMJ>0.7$ radians. Second, we calculate the \ETm\ from charged tracks only, using tracks originating from the primary vertex with \pt$>0.5$\GeV and $\abs{\eta}<2.5$, and require the separation in azimuth from the standard \ETm\ satisfies $\dphiMtkM<0.5$~radians. Third, in the low-\ETm\ region only, we require the \ETm\ significance, defined as the ratio of the \ETm\ and the square root of the scalar sum of transverse energy of all particle-flow objects, to be greater than three.

To reduce the \ttbar\ and WZ backgrounds, events with isolated leptons with \pt$>15$\GeV are rejected.

The \Z boson candidate is defined to be the pair of central ($\abs{\eta}<2.5$) jets, above minimum \pt thresholds given in Table~\ref{tab:zbbh-sel}, that has the greatest vector sum of transverse momenta, \ptjj. Each event is required to pass minimum requirements on \ptjj\ as well as the invariant mass of the jet pair, \mjj.  In the low-\ETm\ category, events with two or more jets in addition to this pair are vetoed.  Each jet in the \Z boson pair are required to be tagged by the CSV algorithm. Separate thresholds are applied to the jets with higher ($\mathrm{CSV}_{\mathrm{max}}$), and lower ($\mathrm{CSV}_{\text{min}}$), values of the CSV discriminator.  The background from V+jets and VV processes is reduced significantly through b tagging, leaving the background in the signal region dominated by sub-processes where the two jets originate from genuine b quarks.

The Z boson mass resolution is improved by roughly 10\% by applying regression techniques similar to those used by the CDF Collaboration~\cite{1107.3026} and in the \VHbb\ search by the CMS Collaboration~\cite{Chatrchyan:2013zna}.  This results in a resolution of approximately 10\%, after all event selection criteria are applied, with a few percent bias on the mass.

\begin{figure}[tbp]
	\begin{center}
		\includegraphics[width=0.48\textwidth]{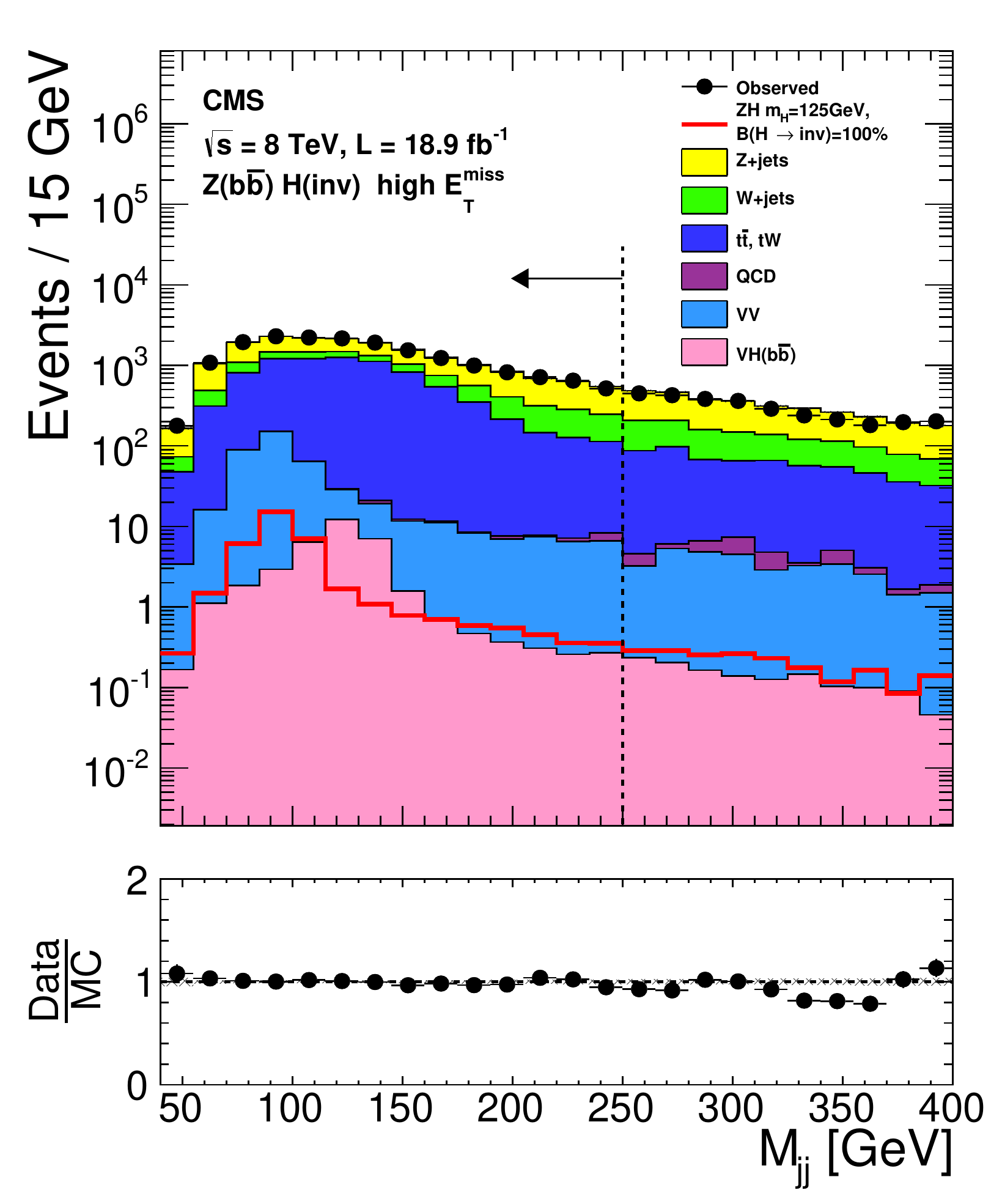}
		\includegraphics[width=0.48\textwidth]{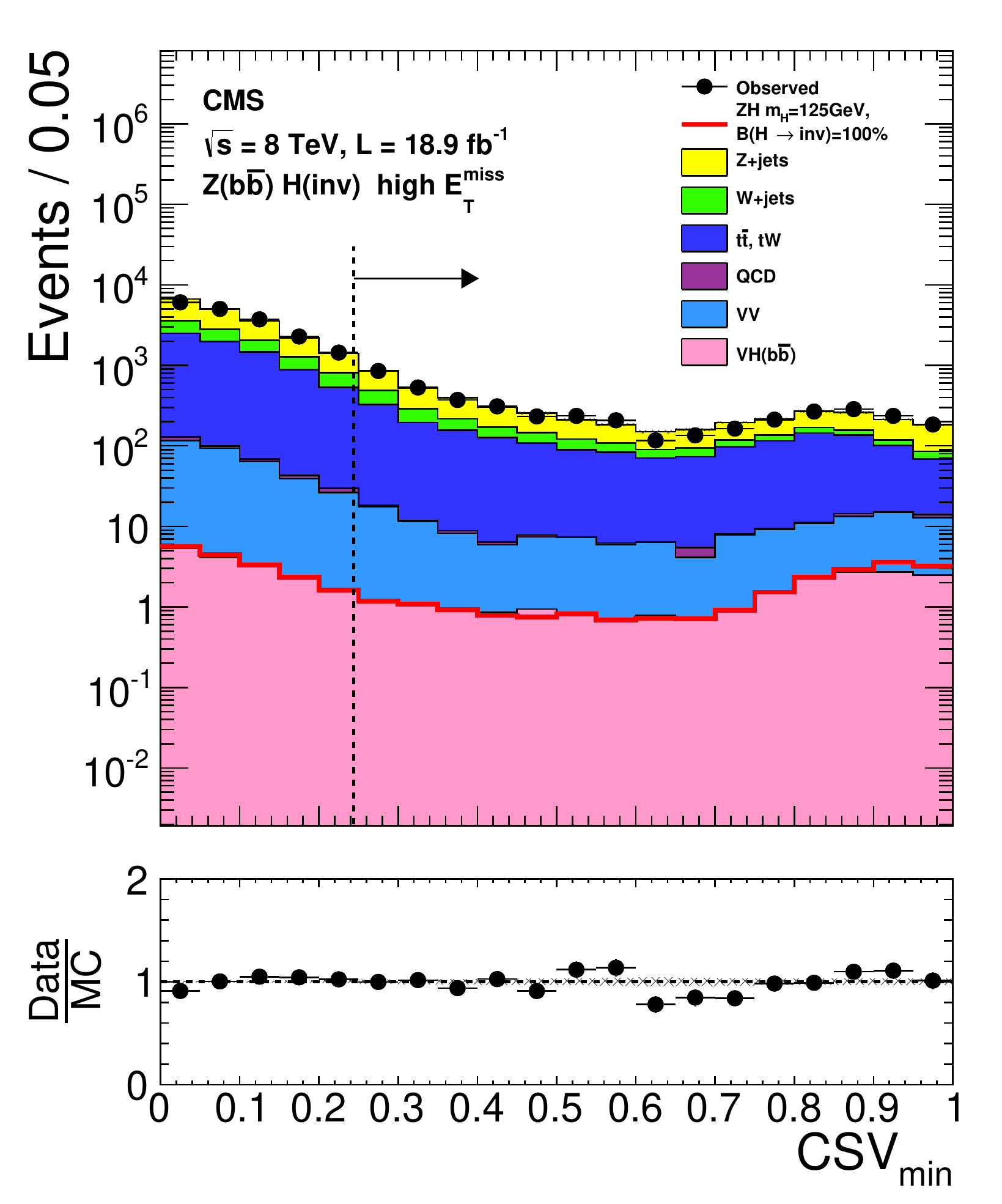}
		\caption{Distributions of \mjj (\cmsLeft) and CSV$_{\mathrm{min}}$ (\cmsRight) in the high-\ETm\ category of the \ZbbHinv\ search, after all other selection requirements.  The simulated background contributions are displayed cumulatively, and the uncertainty in the total background is shown as a hatched region.  The arrows correspond to the cuts applied for the final selection as described in Table~\ref{tab:zbbh-sel}.  The panels below both distributions show the ratio of observed data to expected background events.}
		\label{fig:zbbh_sel}
	\end{center}
\end{figure}

The selection is optimized to give the best signal significance, for a signal with $\mH=125$\GeV and $\BRinv=100$\%.  After all selection criteria, the efficiency for a signal with $\mH=125$\GeV and $\BRinv=100$\% is 4.8\%, while for the most sensitive region of the BDT distribution, defined in Section~\ref{sec:zbbh-syst}, it is 1.75\%.  The effect of the selection on signal and background can be seen in Figure~\ref{fig:zbbh_sel} which shows the \mjj and CSV$_{\mathrm{min}}$ distributions after all other selection requirements.

\begin{table*}[tbp]
	\topcaption{Selection criteria for the \ZbbHinv\ search, in the 3 \ETm regions. The variables used are either described in the text or in Table \ref{tab:bdtvars}.}
	\label{tab:zbbh-sel}
	\centering
		\begin{tabular}{lccc} \hline\hline
			Variable                & \multicolumn{3}{c} {Selection}     \\
									& Low \ETm	& Intermediate \ETm	& High \ETm	\\
			\hline
			\ETm                    & 100--130\GeV 	& 130--170\GeV 	& $>$170\GeV    \\
			\pta            		& $>$60\GeV		& $>$60\GeV		 & $>$60\GeV	 \\
			\ptb            		& $>$30\GeV		& $>$30\GeV		 & $>$30\GeV     \\
			\ptjj                   & $>$100\GeV 		& $>$130\GeV 		& $>$130\GeV    \\
			\mjj                  	& $<$250\GeV		& $<$250\GeV		& $<$250\GeV    \\
			CSV$_{\mathrm{max}}$    & $>$0.679			& $>$0.679			& $>$0.679    \\
			CSV$_{\mathrm{min}}$    & $>$0.244  		& $>$0.244			& $>$0.244     \\
			N additional jets       & $<$2 				& \NA			 & \NA     \\
			N leptons               & $=$0				& $=$0				 & $=$0   \\
			\dphiZH                 & $>$2.0~radians	& $>$2.0~radians	& $>$2.0~radians   \\
			\dphiMJ                 & $>$0.7~radians 	& $>$0.7~radians 	& $>$0.5~radians \\
			\dphiMtkM               & $<$0.5~radians	& $<$0.5~radians	& $<$0.5~radians \\
			\ETm\ significance      & $>$3				& not used			& not used       \\
			\hline\hline
		\end{tabular}
\end{table*}

As mentioned above, a BDT is used in the final stage of the analysis to discriminate signal from backgrounds.  The BDT is trained using simulated samples for signal and all background processes after  the full selection described above. This is performed separately for each Higgs boson mass hypothesis, which cover the range $105 < \mH < 145$~\GeV in 10~\GeV steps. The set of input variables to the BDT is chosen by iterative optimization from a larger number of potentially discriminating variables, and is listed in Table~\ref{tab:bdtvars}.

\begin{table*}[tbp]
	\topcaption{Input variables to the \ZbbHinv\ BDT.}
	\label{tab:bdtvars}
	\centering
		\begin{tabular}{ll}
			\hline\hline
			Variable & \\
			\hline
			\pta, \ptb 	& Transverse momentum of each \Z boson daughter    \\
			\mjj				& Dijet invariant mass                                \\
			\ptjj				& Dijet transverse momentum                          \\
			\ETm				& Missing transverse energy                           \\
			\Naj				& Number of additional jets ($\pt>25\GeV$ and $\abs{\eta} < 4.5$)  \\
			CSV$_{\mathrm{max}}$& Value of CSV for the \Z boson daughter with largest CSV value  \\
			CSV$_{\mathrm{min}}$& Value of CSV for the \Z boson daughter with second largest CSV value \\
			\dphiZH				& Azimuthal angle between \ETm\ and dijet           \\
			\etajj				& Difference in $\eta$ between \Z daughters         \\
			\dRJJ				& Distance in $\eta$-$\phi$ between \Z daughters     \\
			\dThPull			& Color pull angle~\cite{PhysRevLett.105.022001}  \\
			\dphiMJ				& Azimuthal angle between \ETm\ and the closest jet \\
			\AddJetMaxCSV		& Maximum CSV of the additional jets in an event \\
			\AddJetMindR		& Minimum distance between an additional jet and the \Z boson candidate \\
			\mt					& Transverse mass of the ZH system                     \\
			\hline\hline
		\end{tabular}

\end{table*}

\subsection{Background estimation}
\label{sec:zbbh-backgrounds}

All backgrounds are modeled using MC simulation. Control regions in data are used to validate the simulated distributions used as input to the BDT.  These control regions are also used to obtain scale factors to correct the pre-fit normalizations of the dominant \Z+jets, \PW+jets and \ttbar\ backgrounds. We use the same control regions as defined in Ref.~\cite{Chatrchyan:2013zna} for the \ZnnH\ search. For \PW\ backgrounds, the control region is defined using the same kinematic selection as the signal region apart from the lepton veto, which is inverted. For \Z backgrounds we require a mass veto around the Higgs boson mass hypothesis. In addition we split the \Z and \PW\ backgrounds into heavy-flavor enriched regions, by requiring the same b-tag as the signal region, and light-flavor enriched regions, by inverting the b-tag definition of the signal region. For the \ttbar background, the control region is defined by inverting the lepton veto and additional jet criteria, with respect to the signal region definition.

To obtain the scale factors by which the simulated event yields are adjusted, a set of binned likelihood fits are performed to the CSV$_{\mathrm{min}}$ distributions of events in the control regions. These fits are done simultaneously in all control regions, and the normalization of each background process is allowed to vary independently. Fits to several other variables are also performed, to verify consistency. The scale factors account not only for cross section discrepancies, but also residual differences in physics object selection. For the \Z and \PW\ backgrounds, separate sets of scale factors are obtained for each process according to how many of the two jets selected in the \Z boson reconstruction originate from a b quark.  These are labelled: V+udscg for the case where none of the jets originates from a b-quark, V+b for the case where only one of the jets is from a b quark, and V+bb for the case where both jets originate from b quarks. The scale factors obtained are all close to and compatible with unity, except the V+b background where the scale factor is closer to 2, as seen in Ref.~\cite{Chatrchyan:2013zna}.

\begin{figure}[tbp]
  	\begin{center}
    	 \includegraphics[width=0.48\textwidth]{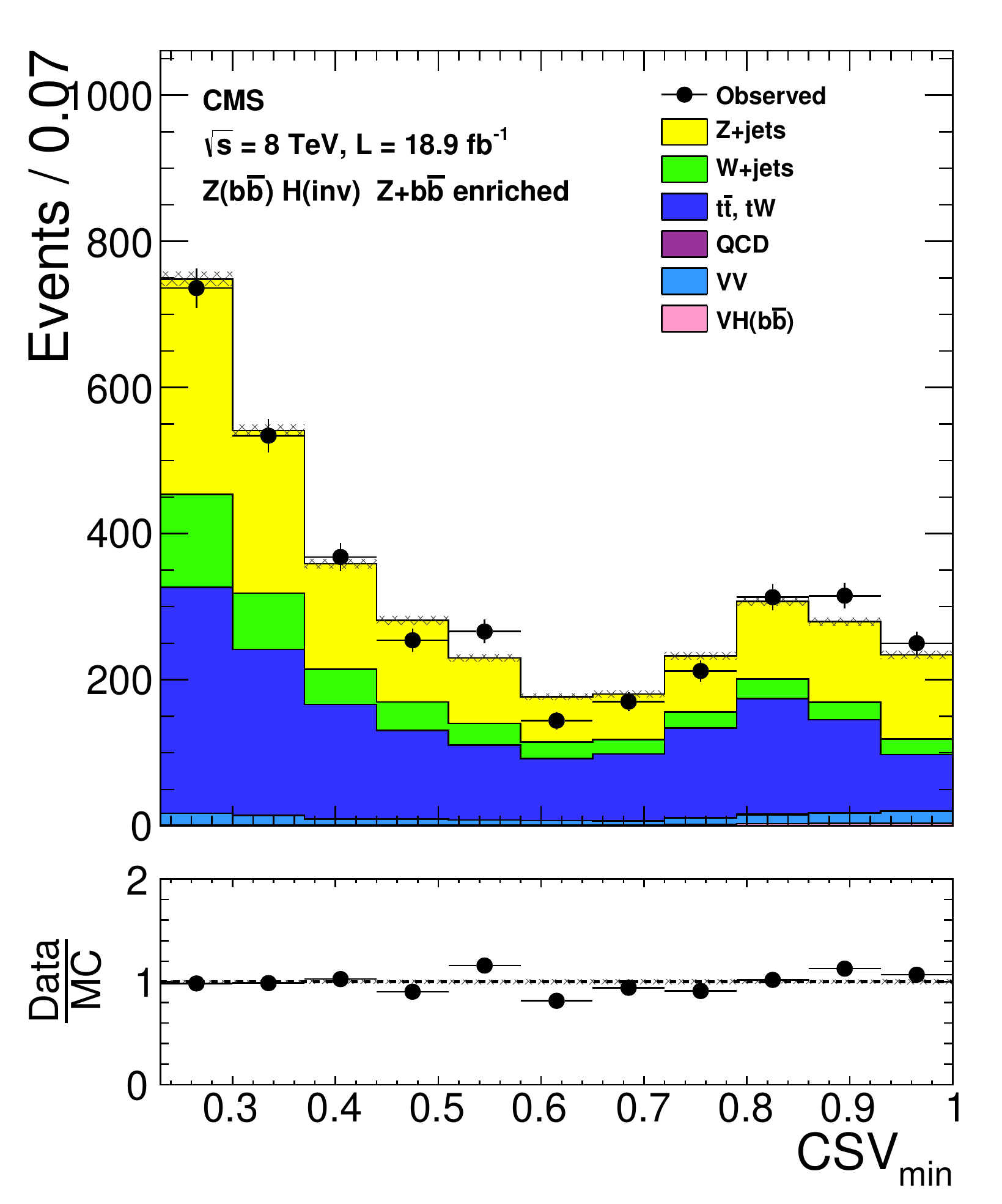}
    	 \includegraphics[width=0.48\textwidth]{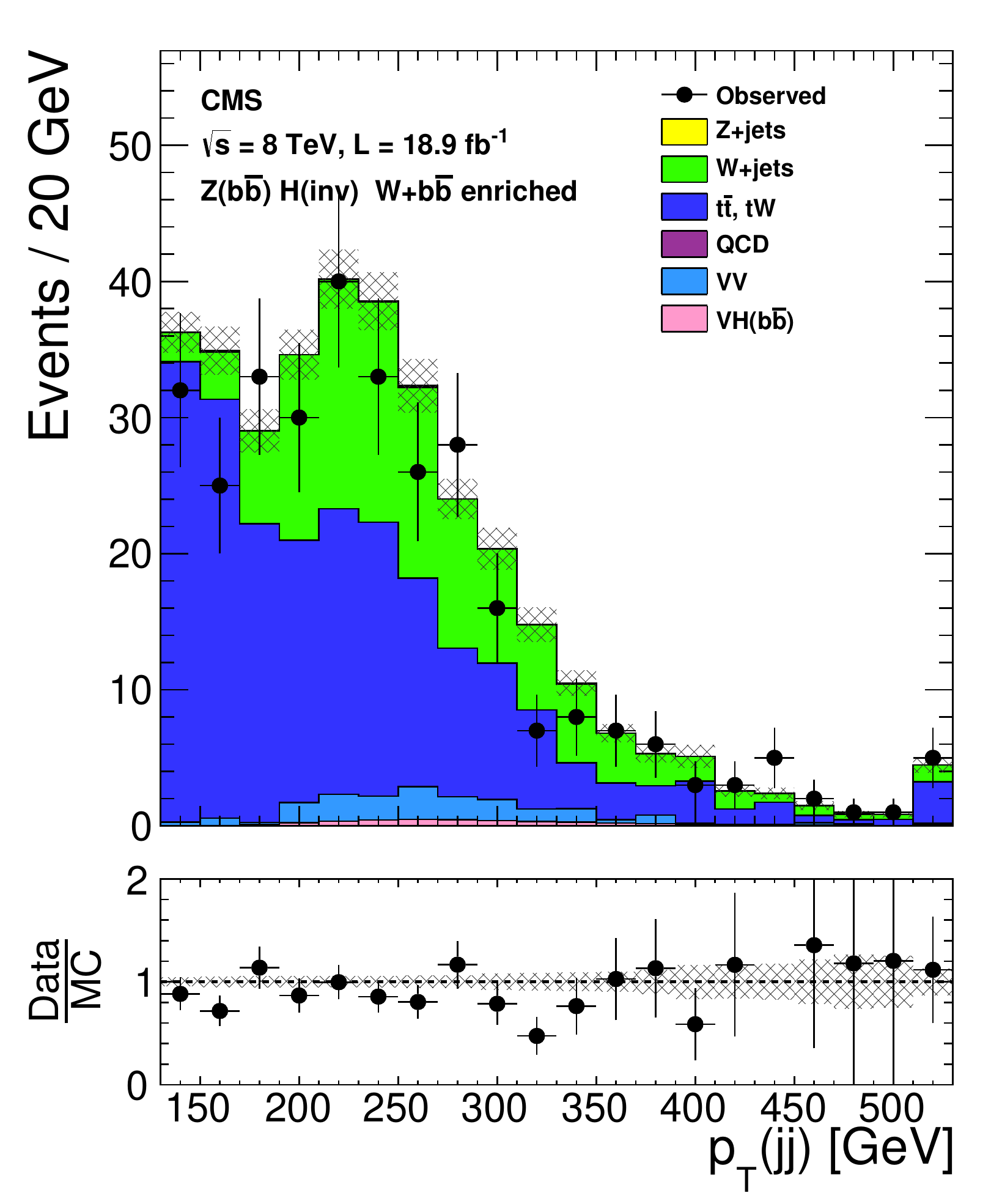}
    	\caption{Distributions in the high-\ETm\ category of the \ZbbHinv\ search: second best CSV among the dijet daughters in the \Zbb\ enriched region (\cmsLeft), and dijet \pt in the \Wbb\ enriched region (\cmsRight).  The simulated background contributions are displayed cumulatively, and the uncertainty in the total background is shown as a hatched region.  The panels below both distributions show the ratio of observed data to expected background events.  An overflow bin is displayed in the right plot.}
    	\label{fig:controlregions}
  	\end{center}
\end{figure}

Table~\ref{tab:bdtyields_rightmost} shows the expected signal and background yields, estimated from MC simulation as described above. Figure~\ref{fig:controlregions} shows the distribution of CSV b-tag discriminant and dijet \pt in the \Zbb\ and \Wbb\ enriched regions, respectively. The high-\ETm\ category is shown, after the data/MC scale factors are applied.

\begin{table*}[htbp]
	\topcaption{Background estimates and signal predictions, together with the observed yields in data, for the most sensitive region in the \ZbbHinv\ BDT analysis.  The signal predictions are given for $\mH=125$\GeV and $\BRinv=100$\%.}
	\label{tab:bdtyields_rightmost}
	\centering
		\begin{tabular}{l|ccc}
			\hline\hline
			Process            &   High \ETm           &   Intermediate \ETm   &   Low \ETm           \\
			\hline
			\ZnnH (SM)         &     2.0 $\pm$   0.3   &     0.4 $\pm$   0.1   &     0.1 $\pm$   0.0  \\
			\WlnH (SM)         &     0.5 $\pm$   0.1   &     0.1 $\pm$   0.0   &     0.1 $\pm$   0.0  \\
			ZZ(bb)             &    27.7 $\pm$   3.1   &    11.6 $\pm$   1.3   &     5.5 $\pm$   0.7  \\
			WZ(bb)             &    10.2 $\pm$   1.6   &     7.3 $\pm$   0.9   &     3.1 $\pm$   0.5  \\
			VV(udscg)          &     5.3 $\pm$   1.1   &     0.3 $\pm$   0.2   &     0.1 $\pm$   0.1  \\
			\Zbb               &    61.8 $\pm$   7.1   &    21.1 $\pm$   2.4   &    13.2 $\pm$   1.6  \\
			\Zb                &    16.7 $\pm$   1.7   &     3.2 $\pm$   1.4   &     0.7 $\pm$   0.9  \\
			\Zudscg            &     7.1 $\pm$   0.3   &     0.6 $\pm$   0.4   &     3.1 $\pm$   2.5  \\
			\Wbb               &    15.8 $\pm$   2.2   &     5.8 $\pm$   0.8   &     3.0 $\pm$   1.4  \\
			\Wb                &     4.7 $\pm$   1.2   &     0.2 $\pm$   0.3   &     0.0 $\pm$   0.0  \\
			\Wudscg            &     4.9 $\pm$   0.2   &     1.1 $\pm$   0.3   &     0.2 $\pm$   0.3  \\
			\ttbar             &    20.4 $\pm$   1.8   &     9.6 $\pm$   1.0   &     8.9 $\pm$   1.1  \\
			Single-top-quark   &     4.1 $\pm$   2.4   &     3.5 $\pm$   2.0   &     2.5 $\pm$   0.7  \\
			QCD                &     0.1 $\pm$   0.1   &     0.0 $\pm$   0.0   &     0.0 $\pm$   0.0  \\
			\hline
			Total backgrounds  &   181.3 $\pm$   9.8   &    64.8 $\pm$   4.1   &    40.5 $\pm$   4.1  \\
			\ZbbHinv           &    12.6 $\pm$   1.1   &     3.6 $\pm$   0.3   &     1.6 $\pm$   0.1  \\
			Observed data      &   204                 &    61                 &    48                \\
			\hline
			S/B           	&     6.9\%               &     5.6\%               &    3.9\%               \\
			\hline\hline
		\end{tabular}
\end{table*}

\subsection{Systematic uncertainty}
\label{sec:zbbh-syst}

Table \ref{tab:syst-ZbbH} lists the uncertainties considered in this channel.  The values quoted are for the most sensitive region of the analysis ($\mathrm{S}/\mathrm{B} >3.5\%$), which corresponds to requirements on the BDT output of $>$0.8, $>$0.7, and $>$0.2 in the low, intermediate, and high-\ETm\ categories, respectively.

Important theoretical uncertainties arise in the signal yield estimation from factorization and renormalization scales, as well as PDF uncertainties, and are estimated as for the \ZllHinv\ and VBF searches.  In addition, uncertainties arising from the QCD NNLO and electroweak NLO corrections discussed in Section~\ref{sec:datasets} are included.

The background estimates are unaffected by theoretical uncertainties, since they are corrected using data/MC scale factors, as discussed in Section~\ref{sec:zbbh-backgrounds}.  However, uncertainties in the background normalization arising from the scale factors themselves are accounted for, by propagating other systematic uncertainties (jet energy scale, jet energy resolution, b tagging efficiency) to the control regions and repeating the fit procedure. Cross section uncertainties of 15\% each are assigned to the single-top-quark backgrounds in the t- and tW-channels, resulting in approximately 1\% uncertainty in the sum of all backgrounds. For the diboson backgrounds, a 7\% cross section uncertainty is assigned, consistent with the CMS measurement of this process~\cite{Chatrchyan:2013oev}, which results in an uncertainty of approximately 4\% in the total background.

As indicated in Table~\ref{tab:syst-ZbbH}, uncertainties affecting the shape of the BDT output are also considered : trigger efficiency, jet energy scale and resolution, unclustered energy, b-tagging efficiency, MC event statistics, lepton momentum scale and pileup.  The jet energy scale and resolution uncertainties are estimated as for the \ZllHinv\ search, resulting in yield uncertainties of 2--4\% and 4--6\%, respectively. The uncertainty associated with b-tagging is taken from uncertainty in the weights applied to MC simulation, mentioned in Section~\ref{sec:reco}. The measured uncertainties for the b-tagging scale factors are: 3\% per b tag, 6\% per charm tag, and 15\% per mistagged jet, originating from gluons and light u, d, s quarks~\cite{Chatrchyan:2012jua}. These translate into yield uncertainties in the 3--5\% range, depending on the channel and the specific process. The shape of the BDT output distribution is also affected by the shape of
the CSV distribution, and is therefore recomputed as the CSV distribution is varied within its uncertainties. The shape uncertainty due to MC modelling of backgrounds is estimated by comparing \MADGRAPH and \HERWIG{}++ results for the V+jets backgrounds, and comparing \MADGRAPH with \POWHEG for \ttbar.

The combined effect of all systematic uncertainties results in a relative increase of about 20\% in the expected upper limit on the \BRinv.

\begin{table*}[tbp]
	\topcaption{Summary of the uncertainties in the \ZbbHinv\ channel. The numbers indicate the change in the total background estimate or in the total signal acceptance when each systematic effect is varied according to its uncertainties. Those uncertainties designated as ``Norm." only affect the normalization of the contributions, while those designated ``Shape" also affect the shapes of the BDT output. In the case of shape variations, the numbers indicate the range of changes across the bins of the distributions. Signal uncertainties are quoted for $\mH=125$\GeV and $\BRinv=100$~\%.  Due to correlations, the total systematic uncertainty is less than the sum in quadrature of the individual uncertainties. The effect is evaluated in the most sensitive region of the BDT output.}
	\label{tab:syst-ZbbH}
	\centering
        \begin{tabular}{llccc}
          \hline \hline
         Type & Source & Background & Signal \\
         & & uncertainty(\%) & uncertainty(\%) \\
  \hline
\multirow{5}{*}{Norm.}                                 		
			& Luminosity                                    &0.9    & 2.6   \\
			& Factorization/renormalization scale and PDFs  &\NA      & 7     \\
			& Signal \pt boost EW/QCD corrections	     	 &\NA     & 6     \\
			& Background data/MC scale factors              &8   & \NA    \\
			& Single-top-quark cross section        		&1    & \NA    \\
			& Diboson cross section                			&4    & \NA         \\
    \hline
\multirow{5}{*}{Shape}
			& Trigger                                       &1    & 5   \\
			& Jet energy scale                              &4    & 3   \\
			& Jet energy resolution                         &3    & 3  \\
			& \ETm\ scale		                     		&1    & 2 \\
			& b tagging                                     &7    & 5    \\
			& MC statistics                        			&3    & 3     \\
			& MC modelling (\VJ\ and \ttbar )               &3    & \NA         \\
			\hline
			&	Total										&	12		&	11		\\
			\hline\hline
		\end{tabular}
\end{table*}

\subsection{Results}
\label{sec:zbbh-results}

\begin{figure*}[htbp]
	\centering
    	 {\includegraphics[width=0.32\textwidth]{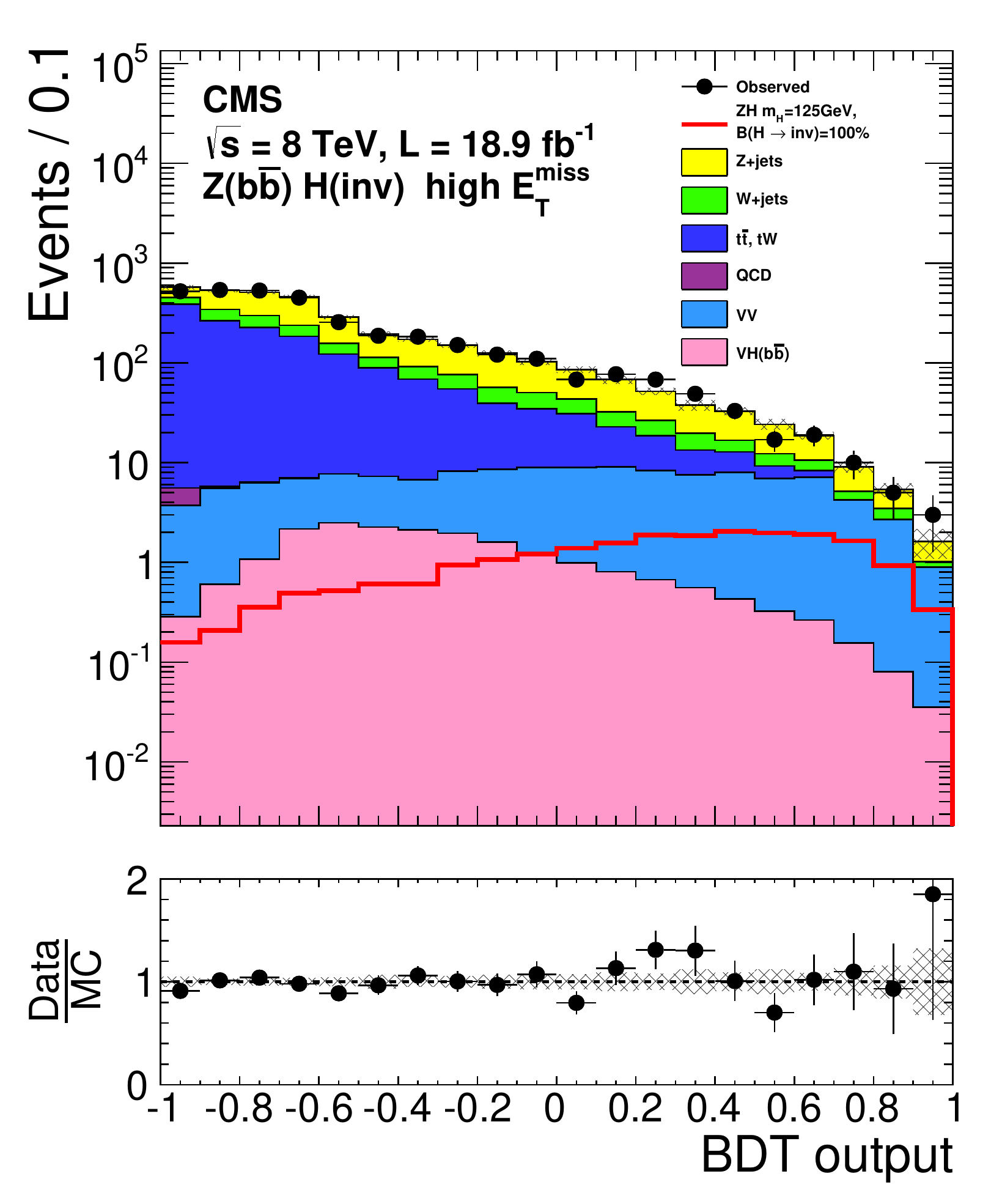}
    	 \includegraphics[width=0.32\textwidth]{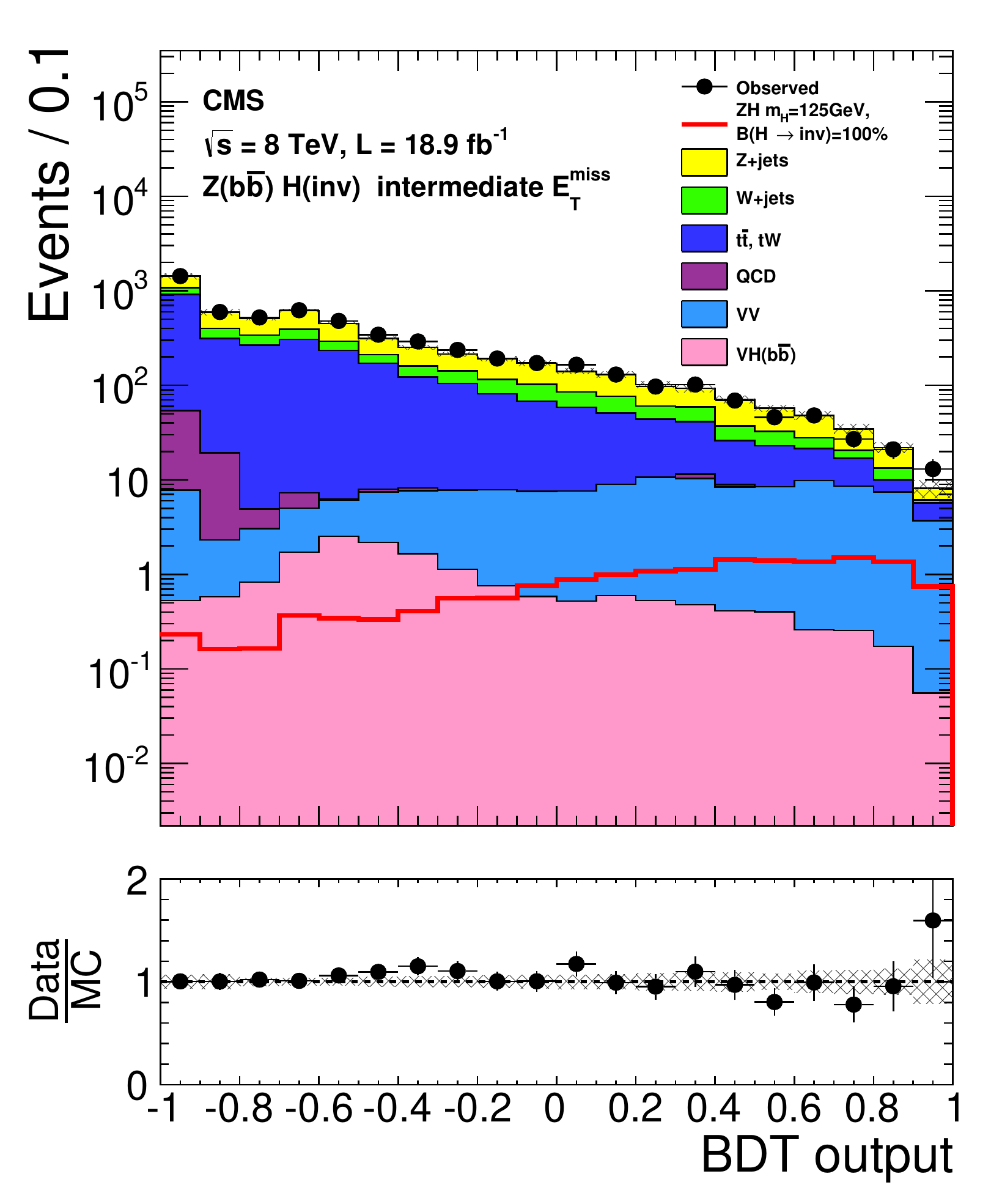}}
    	 \includegraphics[width=0.32\textwidth]{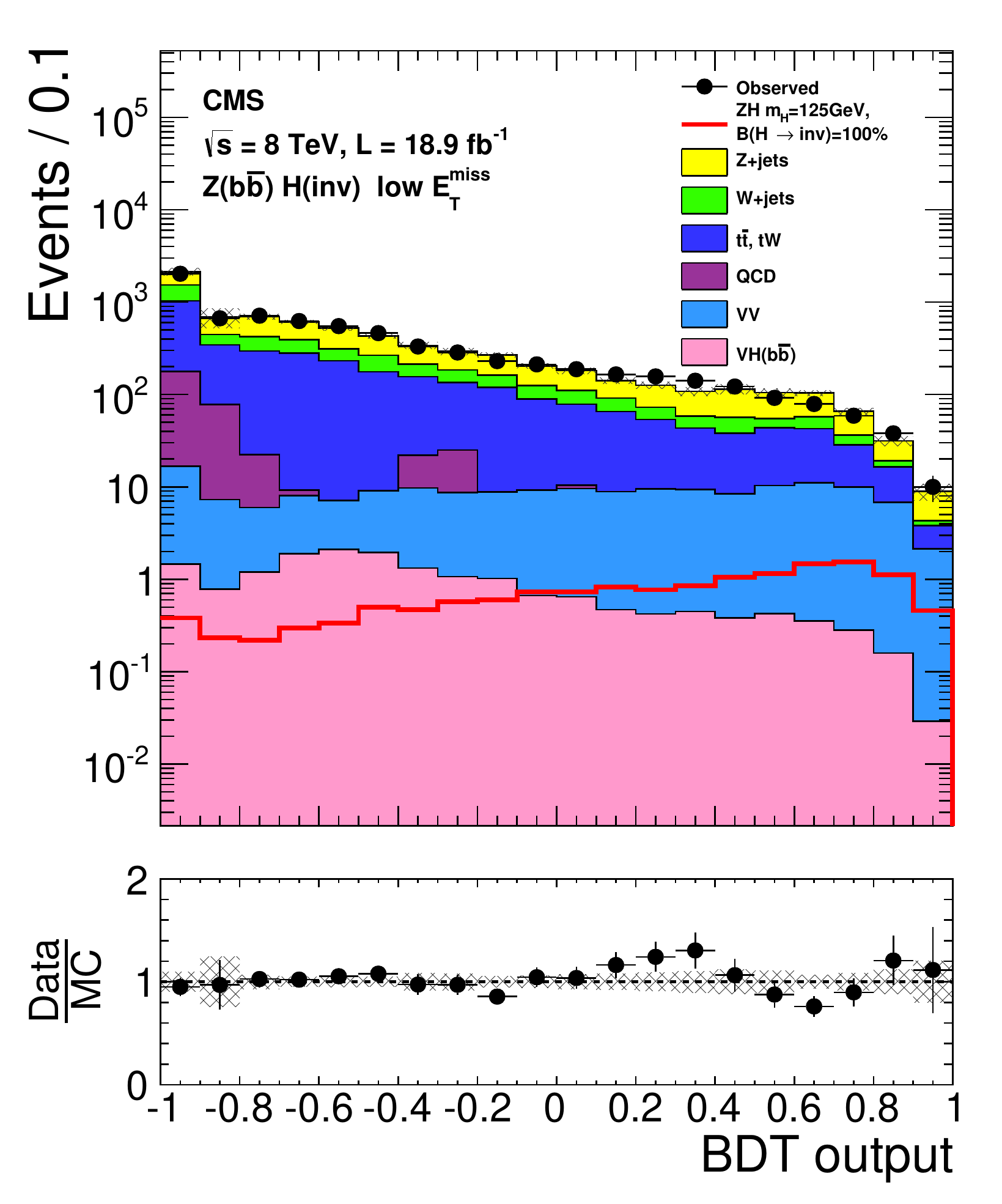}
    	\caption{Distributions of the \ZbbHinv\ BDT output in the high-\ETm\ bin (left), intermediate-\ETm\ bin (center), and low-\ETm\ bin (right) after all selection criteria have been applied.  The simulated background contributions are displayed cumulatively, while a signal corresponding to $\mH=125$\GeV and \BRinv=100\% is superimposed. The uncertainty in the background is shown as a hatched region.  The panels below each distribution show the ratio of observed data to expected background events.  These distributions are used to extract 95\% CL upper limits on the signal.}
    	\label{fig:bdtplots}
\end{figure*}

The number of events observed in data are shown alongside the background estimates in Table~\ref{tab:bdtyields_rightmost}, for the most sensitive regions of the analysis as defined in the previous section.  The BDT output distributions of the three \ETm\ categories are shown in Fig.~\ref{fig:bdtplots}.  In the \ZbbHinv\ search, limits are determined using a fit to the BDT output distribution.  This is performed separately for each Higgs boson mass hypothesis, every 10 GeV in the range 105--145 GeV. In the fit, the shape and normalization for signal and each background component are allowed to vary within the systematic and statistical uncertainties described in Section~\ref{sec:zbbh-syst}. These uncertainties are treated as nuisance parameters in the fit, with appropriate correlations taken into account. All nuisance parameters, including the scale factors described in Section~\ref{sec:zbbh-backgrounds} are adjusted by the fit.

\section{Cross section limits}
\label{sec:limits}

No evidence for a signal is observed in any of the three searches.  We set 95\% CL upper limits on the Higgs boson production cross section times invisible branching fraction, \BRinv, for the VBF and ZH production modes separately.  Limits are calculated using a \CLs\ method~\cite{Read1,junkcls}, based on asymptotic formulae from Ref.~\cite{AsymptoticCLS}, following the standard CMS Higgs boson searches combination technique~\cite{Chatrchyan:2013lba,HiggsCombination}.  Systematic uncertainties are incorporated as nuisance parameters and treated according to the frequentist paradigm described in Ref.~\cite{HiggsCombination}. We also present 95\% CL limits on Higgs boson production cross section times invisible branching fraction normalised to the SM production cross section~\cite{Dittmaier:2011ti,Dittmaier:2012vm}, which we will denote $\xi = \sigma \cdot \BRinv / \sigma_\mathrm{SM}$.  We present limits on $\xi$ for the VBF and ZH modes separately and from the combination of all channels. It should be noted that the assumption of SM production cross sections is an arbitrary choice, as a sizeable invisible width would indicate physics beyond the SM, which may also modify the production cross-section. However, an alternative choice of model for Higgs boson production would essentially scale the limits and provide no further information.

Under the assumption of SM production cross sections and acceptances, we may interpret limits on $\xi$ as limits on the invisible branching fraction of the 125\GeV Higgs boson.

\begin{figure}[htp]
	\centering
		 \includegraphics[width=0.49\textwidth]{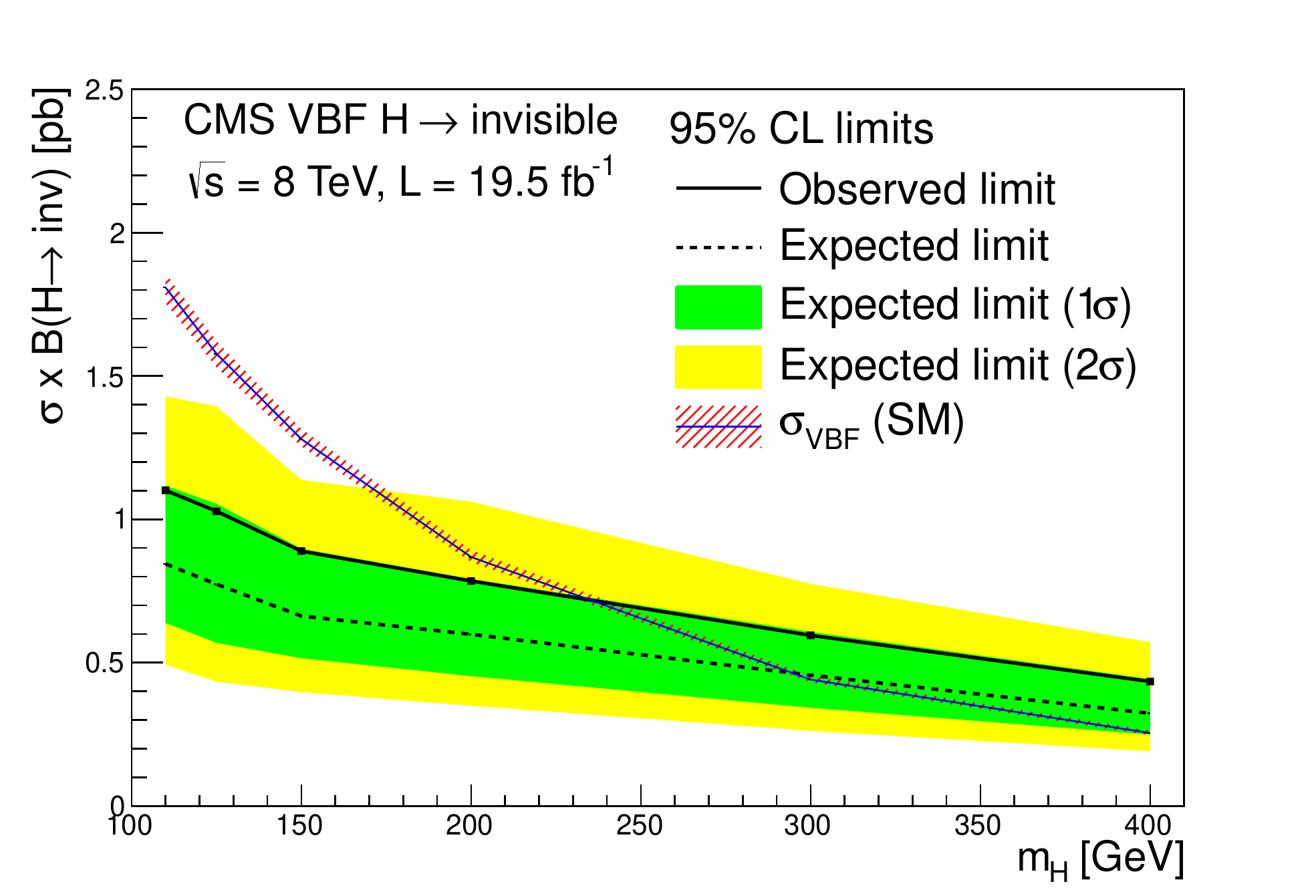}
		 \includegraphics[width=0.49\textwidth]{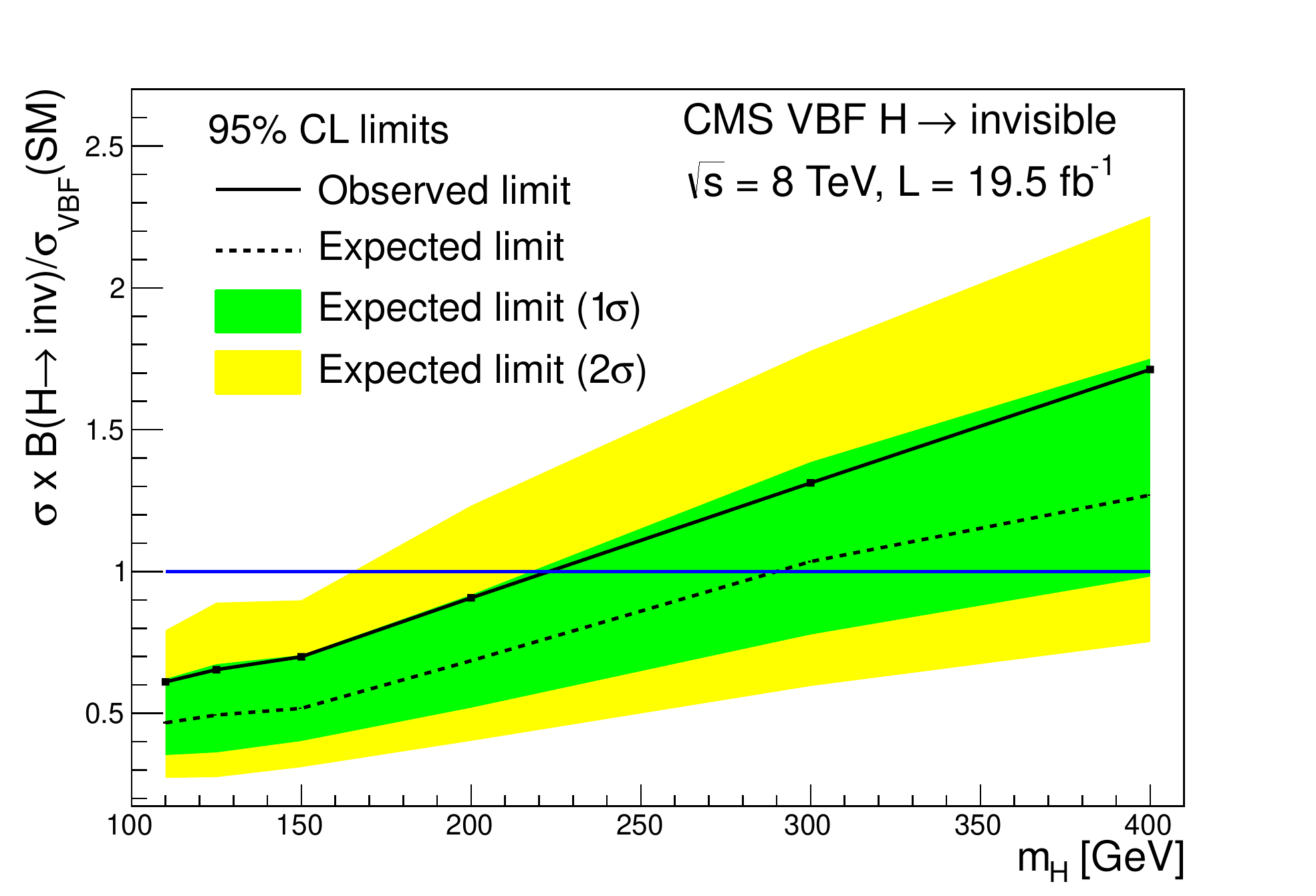}
		\caption{Expected and observed 95\% CL upper limits on the VBF production cross section times invisible branching fraction (\cmsLeft), and normalized to the SM Higgs boson VBF production cross section (\cmsRight).}
		\label{fig:vbfLimit}
\end{figure}

Figure~\ref{fig:vbfLimit} (\cmsLeft) shows the observed and median expected 95\% CL limits on the Higgs boson production cross section times invisible branching fraction, as a function of the Higgs boson mass, for the VBF production mode.  Figure~\ref{fig:vbfLimit} (\cmsRight) shows the corresponding limit on $\xi$.  Assuming the SM VBF production cross section and acceptance, this corresponds to an observed (expected) upper limit on \BRinv\ of 0.65 (0.49) for $\mH=125$\GeV.

\begin{figure}[htbp!]
  \centering
    \includegraphics[width=0.49\textwidth]{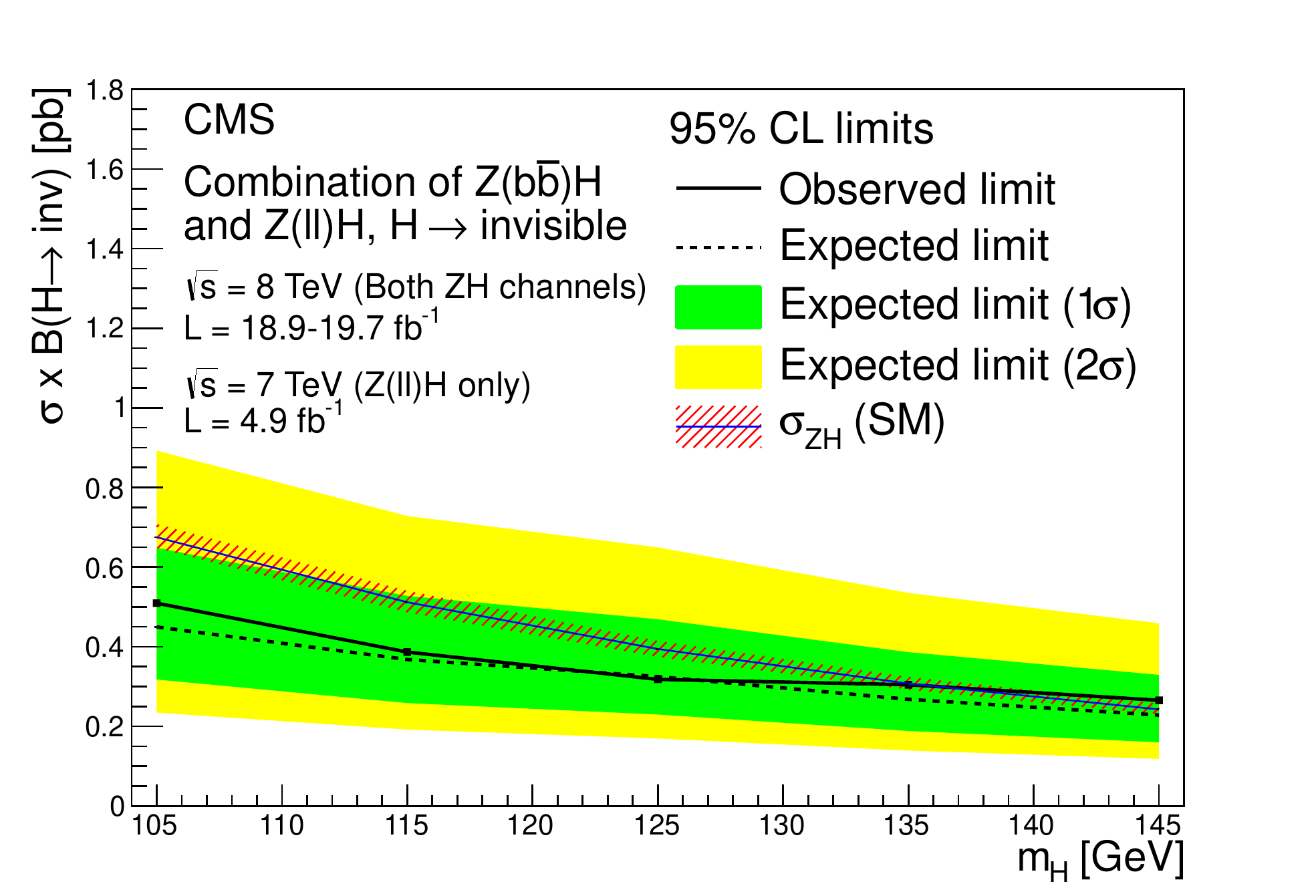}
	 \includegraphics[width=0.49\textwidth]{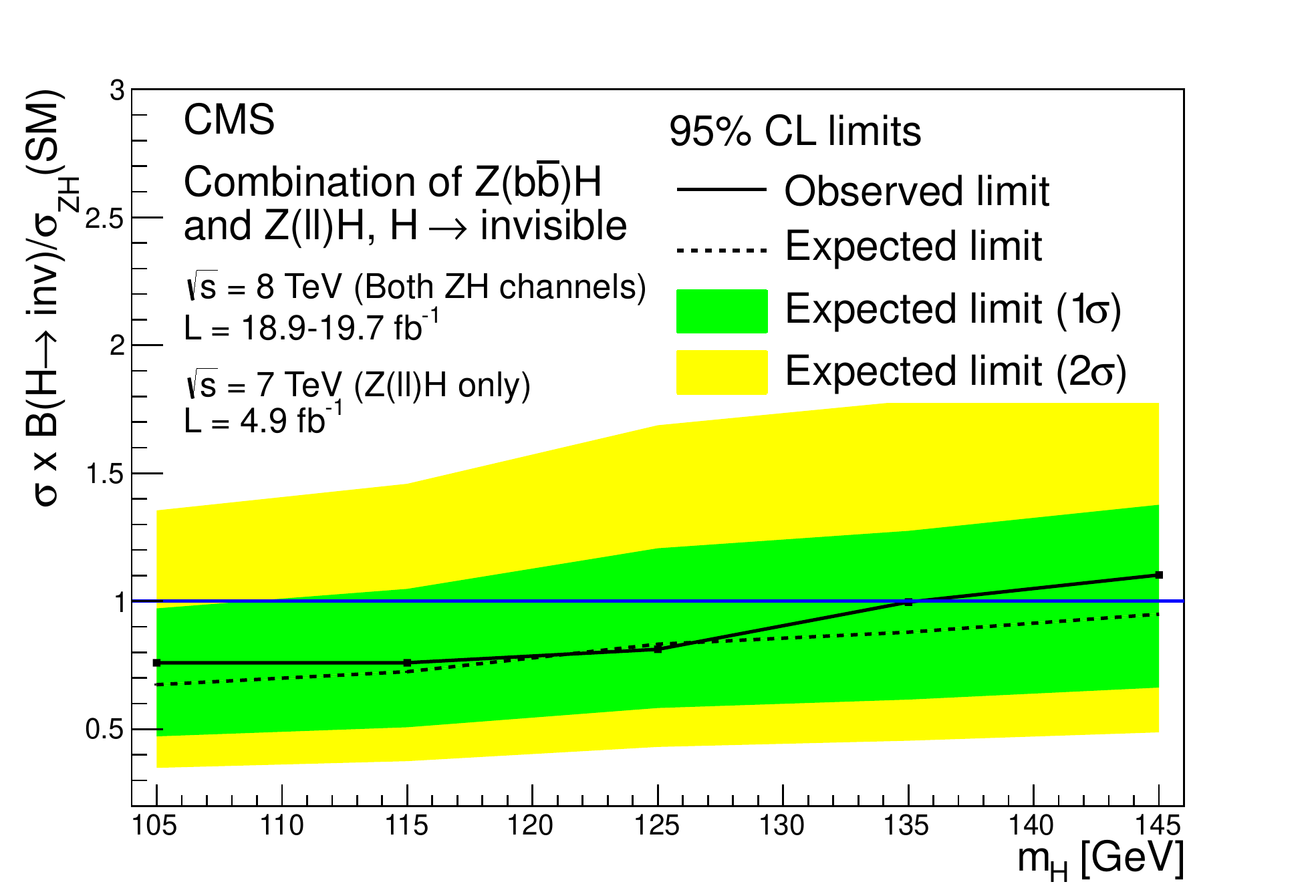}
    \caption{Expected and observed 95\% CL upper limits on the ZH production cross section times invisible branching fraction (\cmsLeft), and normalized to the SM Higgs boson ZH production cross section (\cmsRight).}
    \label{fig:zhLimit}

\end{figure}

The 95\% CL observed and median expected upper limits on the Higgs boson production cross section times invisible branching fraction for the ZH production mode are shown in Fig.~\ref{fig:zhLimit} (\cmsLeft). As for the VBF search, limits on $\xi$ are also shown, in Fig.~\ref{fig:zhLimit} (\cmsRight).  For a Higgs boson with $\mH = 125\GeV$, the observed (expected) upper limit on $\xi$ obtained from the \ZllHinv\ search alone is 0.83 (0.86), and from the \ZbbHinv\ search alone is 1.82 (1.99).  Assuming the SM production cross section and acceptance, we interpret these results as an observed (expected) 95\% CL upper limit on \BRinv\ of 0.81\,(0.83) for $\mH = 125\GeV$.

\begin{figure}[htbp]
  \centering
    \includegraphics[width=\cmsFigWidth]{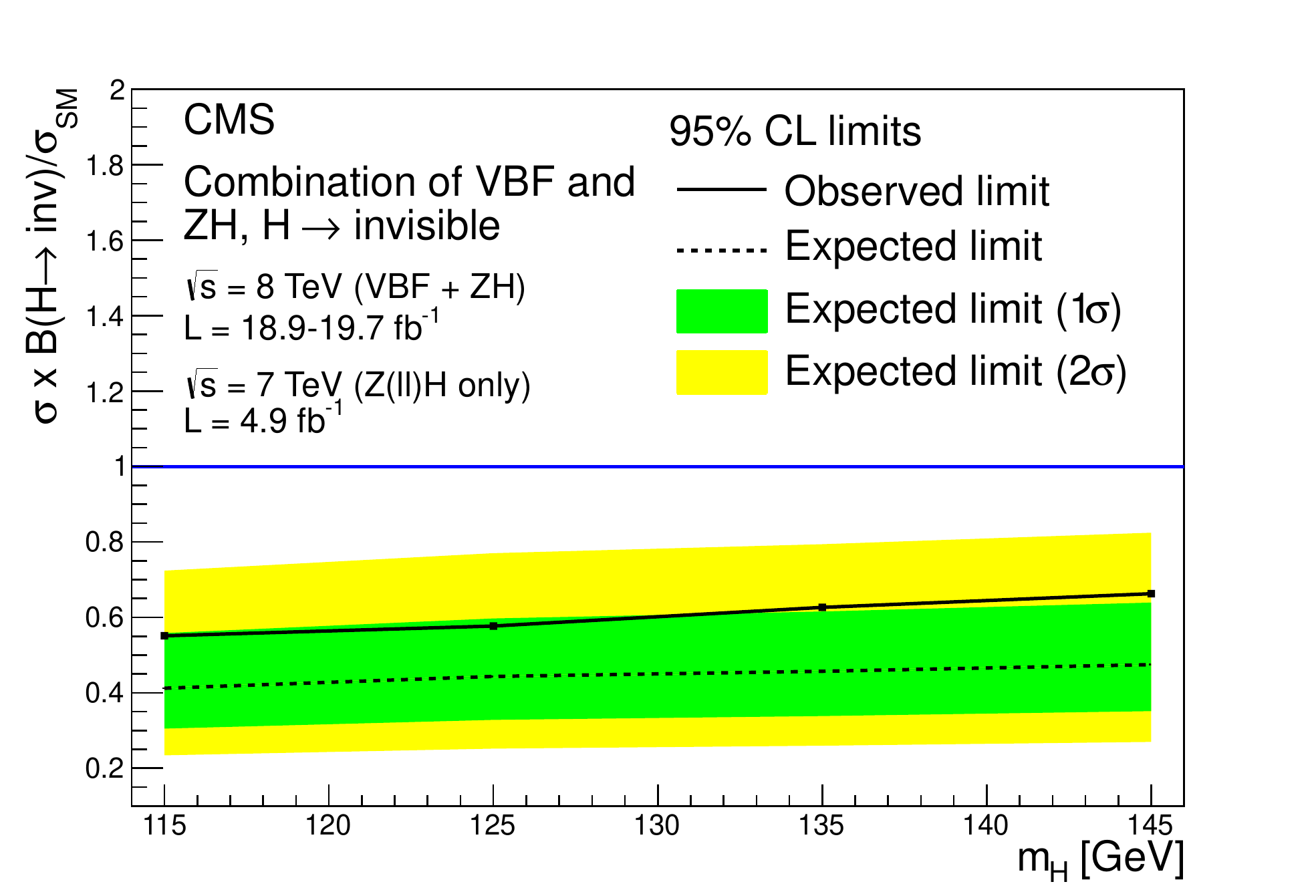}
    \caption{Expected and observed 95\% CL upper limits on $\sigma \cdot \BRinv / \sigma(\mathrm{SM})$.}
    \label{fig:combination}
\end{figure}

By assuming production cross sections as for the SM Higgs boson, the results of the three individual searches may be combined and interpreted as a limit on the invisible branching fraction of the 125\GeV Higgs boson.  The statistical combination fully accounts for correlations between nuisance parameters in the individual searches.  The most important correlated uncertainties are, in decreasing order of importance, the jet energy scale uncertainty, those associated with the signal uncertainty, due to PDF and renormalization/factorization scale variation uncertainties, the total integrated luminosity uncertainty, the lepton momentum scale uncertainties, the jet energy resolution uncertainty and the $\ETm$ energy scale and resolution uncertainties.  The resulting 95\% CL limit on $\xi$ is shown in Fig.~\ref{fig:combination} and summarised in Table~\ref{tab:limits}.  Assuming the SM production cross section and acceptance, the 95\% CL observed upper limit on the invisible branching fraction for $\mH = 125\GeV$ is 0.58, with an expected limit of 0.44.  The corresponding observed (expected) upper limit at 90\% CL is 0.51 (0.38).  These limits significantly improve on the indirect 95\% CL limit of $\BRinv<0.89$ obtained from visible decays~\cite{Chatrchyan:2013lba}.

\begin{table}[h!t]
\centering
\topcaption{Summary of 95\% CL upper limits on $\sigma \cdot \BRinv / \sigma_\mathrm{SM}$ obtained from the VBF search, the combined ZH searches, and the combination of all three searches.}
	\begin{tabular}{lccc}
  		\hline \hline
  		\multirow{3}{*}{$\mH\,(\GeVns{})$} 	& \multicolumn{3}{c}{Observed (expected) upper limits} \\
												& \multicolumn{3}{c}{on $\sigma \cdot  \BRinv / \sigma_\mathrm{SM}$ } 	\\ \cline{2-4}
					& VBF 			& ZH		& VBF+ZH	\\	
  		\hline
  		\hline
  		115			& 0.63 (0.48) 	& 0.76 (0.72) 	& 0.55 (0.41)	\\
  		125			& 0.65 (0.49)	& 0.81 (0.83)	& 0.58 (0.44) 	\\
  		135			& 0.67 (0.50) 	& 1.00 (0.88) 	& 0.63 (0.46)	\\
  		145			& 0.69 (0.51) 	& 1.10 (0.95) 	& 0.66 (0.47)	\\
		200			& 0.91 (0.69)	& \NA			& \NA	\\
		300			& 1.31 (1.04)	& \NA 		& \NA	\\
  		\hline \hline
	\end{tabular}
	\label{tab:limits}
\end{table}

\section{Dark matter interactions}
\label{sec:dm}

We now interpret the experimental upper limit on \BRinv, under the assumption of SM production cross section, in the context of a Higgs-portal model of DM interactions~\cite{Patt:2006fw,Djouadi:2011aa,Djouadi:2012zc}. In these models, a hidden sector can provide viable stable DM particles with direct renormalizable couplings to the Higgs sector of the SM. In direct detection experiments, the elastic interaction between DM and nuclei exchanged through the Higgs boson results in nuclear recoil which can be reinterpreted in terms of DM mass, $M_\chi$, and DM-nucleon cross section.  If the DM candidate has a mass below $\mH/2$, the invisible Higgs boson decay width, \GamInv, can be directly translated to the spin-independent DM-nucleon elastic cross section, as follows for scalar (S), vector (V), and fermionic (f) DM, respectively~\cite{Djouadi:2011aa}:
\begin{equation}
\sigma^\mathrm{SI}_{\mathrm{S}-\mathrm{N}} = \frac{4\GamInv}{\mH^3v^2\beta} \frac{m_\mathrm{N}^4f_\mathrm{N}^2}{(M_\chi+m_\mathrm{N})^2},
\end{equation}
\begin{equation}
\sigma^\mathrm{SI}_{\mathrm{V}-\mathrm{N}} = \frac{16\GamInv M_\chi^4}
{\mH^3 v^2 \beta(\mH^4-4M_\chi^2 \mH^2+12M_\chi^4)}
\frac{m_\mathrm{N}^4f_\mathrm{N}^2}
{(M_\chi+m_N)^2},
\end{equation}
\begin{equation}
\sigma^\mathrm{SI}_{\mathrm{f}-\mathrm{N}} = \frac{8\GamInv M_\chi^2}{\mH^5v^2\beta^3}\frac{m_\mathrm{N}^4f_\mathrm{N}^2}{(M_\chi+m_\mathrm{N})^2}.
\end{equation}
Here, $m_\mathrm{N}$ represents the nucleon mass, taken as the average of proton and neutron masses, 0.939\GeV, while $\sqrt{2}v$ is the Higgs vacuum expectation value of 246\GeV, and $\beta=\sqrt{1-4M^2_\chi/{\mH}^2}$. The dimensionless quantity $f_\mathrm{N}$~\cite{Djouadi:2011aa} parameterizes the Higgs-nucleon coupling; we take the central values of $f_\mathrm{N}=0.326$ from a lattice calculation~\cite{PhysRevD.81.014503}, while we use results from the MILC Collaboration~\cite{PhysRevLett.103.122002} for the minimum (0.260) and maximum (0.629) values.  We convert the invisible branching fraction to the invisible width using $\BRinv =  \GamInv / (\Gamma_\mathrm{SM} + \GamInv)$, where $\Gamma_\mathrm{SM}= 4.07$\MeV.

\begin{figure}[htbp]
	\centering
  		 \includegraphics[width=\cmsFigWidth]{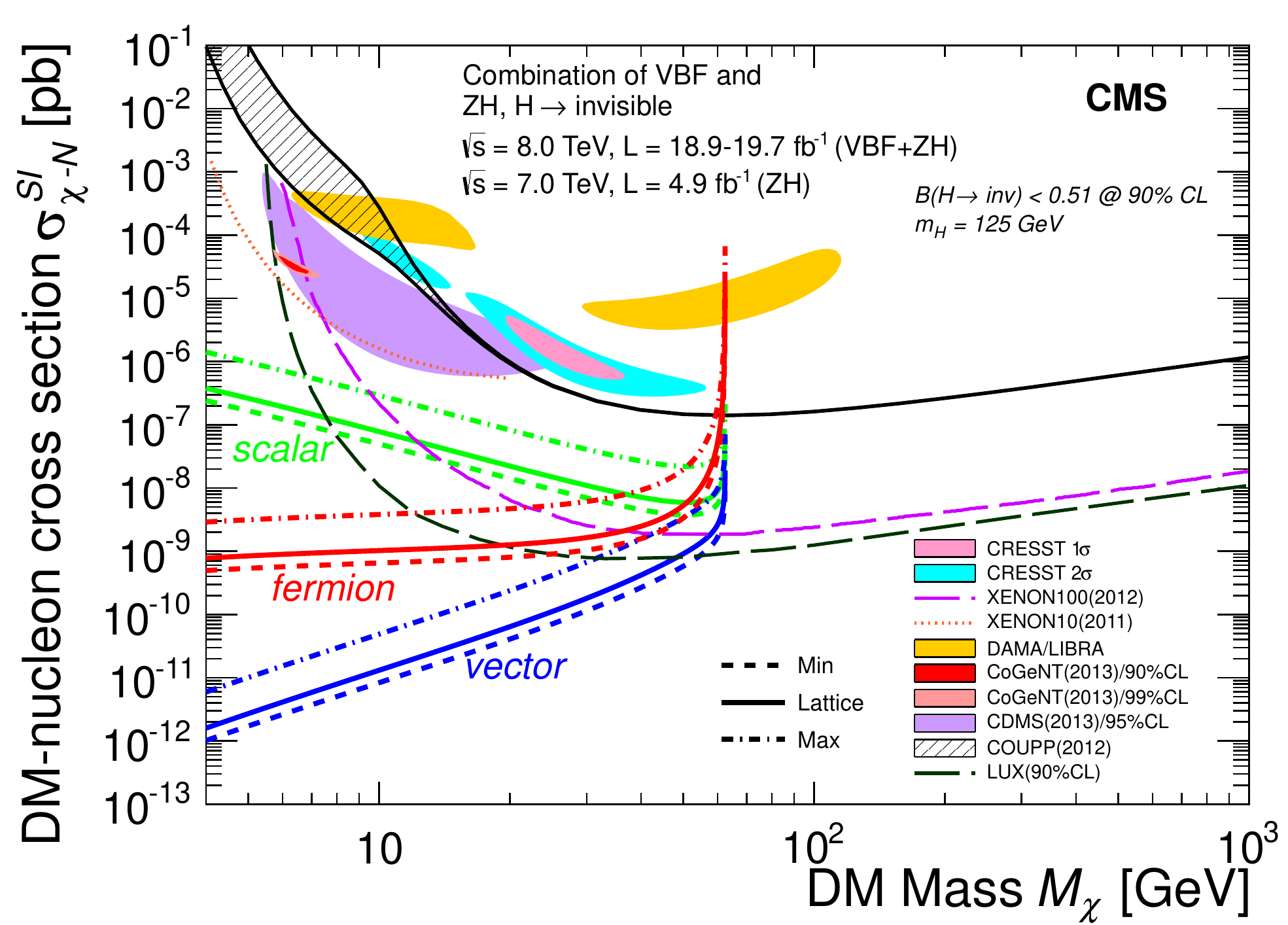}
  		\caption{Upper limits on the spin-independent DM-nucleon cross section $\sigma^\mathrm{SI}_{\chi-\mathrm{N}}$ in Higgs-portal models, derived for $\mH = 125\GeV$ and $\BRinv < 0.51$ at 90\% CL, as a function of the DM mass. Limits are shown separately for scalar, vector and fermion DM. The solid lines represent the central value of the Higgs-nucleon coupling, which enters as a parameter, and is taken from a lattice calculation, while the dashed and dot-dashed lines represent lower and upper bounds on this parameter.  Other experimental results are shown for comparison, from the CRESST~\cite{Angloher:2011uu}, XENON10~\cite{PhysRevLett.107.051301}, XENON100~\cite{Aprile:2012nq}, DAMA/LIBRA~\cite{Bernabei:2008yi,Savage:2008er}, CoGeNT~\cite{Aalseth:2012if}, CDMS II~\cite{Agnese:2013rvf}, COUPP~\cite{Behnke:2012ys}, LUX~\cite{Akerib:2013tjd} Collaborations.}
		\label{fig:DM}
\end{figure}

Figure~\ref{fig:DM} shows upper limits at 90\% CL on the DM-nucleon cross section as a function of the DM mass, derived from the experimental upper limit on \BRinv\ for $\mH=125$\GeV, in the scenarios where the DM candidate is a scalar, a vector, or a Majorana fermion.

\section{Summary}
\label{sec:conclusion}

A search for invisible decays of Higgs bosons has been performed, using the vector boson fusion and associated ZH production modes, with $\Z\to\ell\ell$ or $\Z\to \bbbar$.  No evidence for a signal is observed in any channel.  Using a \CLs method, upper limits are placed on the Higgs boson production cross section times invisible branching fraction, for the VBF and ZH channels separately and combined. These results improve the exclusion in terms of $\sigma \cdot \BRinv / \sigma_\mathrm{SM}$ for $\mH>113$\GeV with respect to the limits obtained at LEP~\cite{Searches:2001ab}.  By assuming standard model production cross sections, and combining all channels, the upper limit on the invisible branching fraction of a Higgs boson for $\mH=125$\GeV, is found to be 0.58, with an expected limit of 0.44, at 95\% confidence level.  These limits assume the signal acceptance of a SM Higgs boson.  These constraints are more stringent than the indirect limits obtained from visible Higgs boson decays.  Finally, the result is interpreted in a Higgs-portal model of dark matter~\cite{Djouadi:2012zc}.  Strong limits, beyond those from direct searches, are obtained on the dark matter nucleon cross section for light dark matter.

\section*{Acknowledgements}
{\tolerance=900\hyphenation{Bundes-ministerium Forschungs-gemeinschaft Forschungs-zentren} We congratulate our colleagues in the CERN accelerator departments for the excellent performance of the LHC and thank the technical and administrative staffs at CERN and at other CMS institutes for their contributions to the success of the CMS effort. In addition, we gratefully acknowledge the computing centres and personnel of the Worldwide LHC Computing Grid for delivering so effectively the computing infrastructure essential to our analyses. Finally, we acknowledge the enduring support for the construction and operation of the LHC and the CMS detector provided by the following funding agencies: the Austrian Federal Ministry of Science and Research and the Austrian Science Fund; the Belgian Fonds de la Recherche Scientifique, and Fonds voor Wetenschappelijk Onderzoek; the Brazilian Funding Agencies (CNPq, CAPES, FAPERJ, and FAPESP); the Bulgarian Ministry of Education and Science; CERN; the Chinese Academy of Sciences, Ministry of Science and Technology, and National Natural Science Foundation of China; the Colombian Funding Agency (COLCIENCIAS); the Croatian Ministry of Science, Education and Sport, and the Croatian Science Foundation; the Research Promotion Foundation, Cyprus; the Ministry of Education and Research, Recurrent financing contract SF0690030s09 and European Regional Development Fund, Estonia; the Academy of Finland, Finnish Ministry of Education and Culture, and Helsinki Institute of Physics; the Institut National de Physique Nucl\'eaire et de Physique des Particules~/~CNRS, and Commissariat \`a l'\'Energie Atomique et aux \'Energies Alternatives~/~CEA, France; the Bundesministerium f\"ur Bildung und Forschung, Deutsche Forschungsgemeinschaft, and Helmholtz-Gemeinschaft Deutscher Forschungszentren, Germany; the General Secretariat for Research and Technology, Greece; the National Scientific Research Foundation, and National Innovation Office, Hungary; the Department of Atomic Energy and the Department of Science and Technology, India; the Institute for Studies in Theoretical Physics and Mathematics, Iran; the Science Foundation, Ireland; the Istituto Nazionale di Fisica Nucleare, Italy; the Korean Ministry of Education, Science and Technology and the World Class University program of NRF, Republic of Korea; the Lithuanian Academy of Sciences; the Ministry of Education, and University of Malaya (Malaysia); the Mexican Funding Agencies (CINVESTAV, CONACYT, SEP, and UASLP-FAI); the Ministry of Business, Innovation and Employment, New Zealand; the Pakistan Atomic Energy Commission; the Ministry of Science and Higher Education and the National Science Centre, Poland; the Funda\c{c}\~ao para a Ci\^encia e a Tecnologia, Portugal; JINR, Dubna; the Ministry of Education and Science of the Russian Federation, the Federal Agency of Atomic Energy of the Russian Federation, Russian Academy of Sciences, and the Russian Foundation for Basic Research; the Ministry of Education, Science and Technological Development of Serbia; the Secretar\'{\i}a de Estado de Investigaci\'on, Desarrollo e Innovaci\'on and Programa Consolider-Ingenio 2010, Spain; the Swiss Funding Agencies (ETH Board, ETH Zurich, PSI, SNF, UniZH, Canton Zurich, and SER); the National Science Council, Taipei; the Thailand Center of Excellence in Physics, the Institute for the Promotion of Teaching Science and Technology of Thailand, Special Task Force for Activating Research and the National Science and Technology Development Agency of Thailand; the Scientific and Technical Research Council of Turkey, and Turkish Atomic Energy Authority; the National Academy of Sciences of Ukraine, and State Fund for Fundamental Researches, Ukraine; the Science and Technology Facilities Council, UK; the US Department of Energy, and the US National Science Foundation.

Individuals have received support from the Marie-Curie programme and the European Research Council and EPLANET (European Union); the Leventis Foundation; the A. P. Sloan Foundation; the Alexander von Humboldt Foundation; the Belgian Federal Science Policy Office; the Fonds pour la Formation \`a la Recherche dans l'Industrie et dans l'Agriculture (FRIA-Belgium); the Agentschap voor Innovatie door Wetenschap en Technologie (IWT-Belgium); the Ministry of Education, Youth and Sports (MEYS) of Czech Republic; the Council of Science and Industrial Research, India; the Compagnia di San Paolo (Torino); the HOMING PLUS programme of Foundation for Polish Science, cofinanced by EU, Regional Development Fund; and the Thalis and Aristeia programmes cofinanced by EU-ESF and the Greek NSRF; and the royal patronage of H.R.H. Princess Maha Chakri
Sirindhorn of Thailand.

\par}
\bibliography{auto_generated}   

\cleardoublepage \appendix\section{The CMS Collaboration \label{app:collab}}\begin{sloppypar}\hyphenpenalty=5000\widowpenalty=500\clubpenalty=5000\textbf{Yerevan Physics Institute,  Yerevan,  Armenia}\\*[0pt]
S.~Chatrchyan, V.~Khachatryan, A.M.~Sirunyan, A.~Tumasyan
\vskip\cmsinstskip
\textbf{Institut f\"{u}r Hochenergiephysik der OeAW,  Wien,  Austria}\\*[0pt]
W.~Adam, T.~Bergauer, M.~Dragicevic, J.~Er\"{o}, C.~Fabjan\cmsAuthorMark{1}, M.~Friedl, R.~Fr\"{u}hwirth\cmsAuthorMark{1}, V.M.~Ghete, C.~Hartl, N.~H\"{o}rmann, J.~Hrubec, M.~Jeitler\cmsAuthorMark{1}, W.~Kiesenhofer, V.~Kn\"{u}nz, M.~Krammer\cmsAuthorMark{1}, I.~Kr\"{a}tschmer, D.~Liko, I.~Mikulec, D.~Rabady\cmsAuthorMark{2}, B.~Rahbaran, H.~Rohringer, R.~Sch\"{o}fbeck, J.~Strauss, A.~Taurok, W.~Treberer-Treberspurg, W.~Waltenberger, C.-E.~Wulz\cmsAuthorMark{1}
% ...
\vskip\cmsinstskip
% ...
1:~~Also at Vienna University of Technology, Vienna, Austria\\
2:~~Also at CERN, European Organization for Nuclear Research, Geneva, Switzerland\\
% ...

\end{sloppypar}
\end{document}